\documentclass[journal]{IEEEtran}

\usepackage{amssymb}
\usepackage[utf8]{inputenc}
\usepackage[T1]{fontenc}
\usepackage{array,multirow}
\usepackage{rotating}
\usepackage{color}
\usepackage{arydshln}
\usepackage[numbers,sort&compress]{natbib}
\usepackage{url}
\newcommand*\rot{\rotatebox{90}}
\newcolumntype{P}[1]{>{\centering\arraybackslash}p{#1}}
\usepackage[]{units}

\usepackage{mathtools}

\DeclarePairedDelimiter\floor{\lfloor}{\rfloor}

\usepackage{algorithm}

\usepackage{algpseudocode}
\usepackage{varwidth}

\IEEEoverridecommandlockouts                              

\title{In-ear EEG biometrics for feasible and readily collectable real-world person authentication}

\author{Takashi Nakamura, Valentin Goverdovsky and Danilo P. Mandic

\thanks{Takashi Nakamura, Valentin Goverdovsky and Danilo P. Mandic are with Department of Electrical and Electronic Engineering, Imperial College London,
London, SW7 2AZ, United Kingdom, \{takashi.nakamura14, goverdovsky, d.mandic\}@imperial.ac.uk}%
}

\begin{document}
\maketitle
\thispagestyle{empty}
\pagestyle{empty}

\begin{abstract}
The use of EEG as a biometrics modality has been investigated for about a decade, however its feasibility in real-world applications is not yet conclusively established, mainly due to the issues with collectability and reproducibility. To this end, we propose a readily deployable EEG biometrics system based on a `one-fits-all'  viscoelastic generic in-ear EEG sensor (collectability), which does not require skilled assistance or cumbersome preparation. 
Unlike most existing studies, we consider data recorded over multiple recording days and for multiple subjects (reproducibility) while, for rigour, the training and test segments are not taken from the same recording days.
A robust approach is considered based on the resting state with eyes closed paradigm, the use of both parametric (autoregressive model) and non-parametric (spectral) features, and supported by simple and fast cosine distance, linear discriminant analysis and support vector machine classifiers. 
Both the verification and identification forensics scenarios are considered and the achieved results are on par with the studies based on impractical on-scalp recordings.
Comprehensive analysis over a number of subjects, setups, and analysis features demonstrates the feasibility of the proposed ear-EEG biometrics, and its potential in resolving the critical collectability, robustness, and reproducibility issues associated with current EEG biometrics. 
\end{abstract}
\section{Introduction}
Person authentication refers to the process of confirming the claimed identity of an individual, and is already present in many aspects of life, such as electronic banking and border control. The existing authentication strategies can be categorised into: 1) knowledge-based (password, PIN), 2) token-based (passport, card), 3) biometric (fingerprints, iris) \cite{Jain2000}.
Most extensively used recognition methods are based on knowledge and tokens, however, these are also most vulnerable to fraud, such as theft and forgery, and can be straightforwardly used by imposters.
In contrast, biometric recognition methods rest upon unique physiological or behavioural characteristics of a person, which then serve as `biomarkers' of an individual, and thus largely overcome the above vulnerabilities. However, at present, biometric authentication systems are cumbersome to administer and require considerable computational and man-power overloads, such as special recording devices and the corresponding classification software.

With the current issues in global security, we have witnessed a rapid growth in biometrics applications based on various modalities, which include palm patterns with high spectral wave \cite{Sato2013}, patterns of eye movement \cite{Holland2013}, patterns in the electrocardiogram (ECG) \cite{Odinaka2010}, and otoacoustic emissions \cite{Liu2014}.
Each such biometric modality has its strengths and weaknesses, and typically suits only a chosen type of application and its corresponding scenarios \cite{Prabhakar2003}. A robust biometric system in the real-world should satisfy the following requirements \cite{Jain2000}: 
\begin{itemize}
	\item \emph{Universality}: each person should possess the given biometric characteristic,
	\item \emph{Uniqueness}: not any two people should share the given characteristic,
	\item \emph{Permanence}: the biometric characteristic should neither change with time nor be alterable,
	\item \emph{Collectability}: the characteristic should be readily measurable by a sensor and readily quantifiable.
\end{itemize}
In addition, a practical biometric system must be harmless to the users, and should maximise the trade-off between performance, acceptability, and circumvention; in other words, it should be designed with the accuracy, speed, and resource requirements in mind \cite{Prabhakar2003}. 

One of the currently investigated biometric modalities is the electroencephalogram (EEG), an electrical potential between specific locations on the scalp which arises as a result of the electrical field generated by assemblies of cortical neurons, and reflects brain activity of an individual, such as intent \cite{Wolpaw2002}. From a biometrics perspective, the EEG fulfils the above requirements of \emph{universality}, as it can be recorded from anyone, together with the \emph{uniqueness}. {\color{black} Specifically, the individual differences of EEG alpha rhythms has been examined \cite{Johnson1959} and reported to exhibit a significant power in discriminating individuals \cite{Berkhout1968} in the area of clinical neurophysiology. The brain activity is neither exposed to surroundings nor possible to be captured at a distance, therefore the brain patterns of an individual are robust to forgery, unlike face, iris, and fingerprints.} 
The EEG is therefore more robust against imposters' attacks than other biometrics and among different technologies to monitor brain function.
However, in order to utilise EEG signals in the real-world, several key properties such as \emph{permanence} and \emph{collectability} must be further addressed.

The `proof-of-concept' for EEG biometrics, was introduced in our own previous works in \cite{Palaniappan2007a, Palaniappan2007}, and most of the follow-up studies were conducted over only one day (or even over one single trial) recording with EEG channels covering the entire head, while in the classification stage, the training and validation datasets were randomly selected from the same recording day (or the same trial). Apart from its usefulness as a proof-of-concept, this setup does not satisfy the feasibility requirement for a real-world biometric application, since:  
\begin{itemize}
	\item Recording scalp-EEG with multiple electrodes is time-consuming to set-up and cumbersome to wear. Such a sensor therefore does not meet \emph{collectablity} requirement. 
	\item EEG recordings from one day (or a single trial) cannot truly evaluate the performance in identifying features of an individual, as this scenario does not satisfy the \emph{permanence} requirement either, see details in Section \ref{sec:feasible_application}.  
	\item The training and validation data within this scenario are inevitably mixed, thereby introducing a \emph{performance bias} in classification. The classification results from such studies are therefore unrealistically high, and we shall refer to this setting as the \emph{\textbf{biased scenario}}.
\end{itemize}
Therefore, for feasible EEG biometrics, the EEG sensor should be wearable, easy to administer, and fast to set-up, while in order to evaluate the performance, the recorded signal should be split in a rigorous way -- the training and validation datasets in the classification stage should be created so as \emph{not to share the same recording days}, a setting we refer to as the \emph{\textbf{rigorous scenario}}. While a considerable body of research has been undertaken to explore the EEG biometrics and to find the most informative subject-specific characteristic of EEG (\emph{uniqueness}), most studies either focused on reducing the number of electrodes (\emph{collectability}) or on evaluating whether the traits are temporally robust (\emph{permanence}) by using EEG data obtained over multiple recording days; for more details see Section \ref{sec:previous_protocol}. 

In this paper, based on our works in \cite{Palaniappan2007} and \cite{Looney2012}, we bring EEG-based biometrics into the real-world by resolving the following critical issues:
\begin{enumerate}
	\item \emph{Collectability}. Biometrics verification is evaluated with a wearable and easy to set-up in-ear sensor, the so-called ear-EEG \cite{Looney2012}, 
	\item \emph{Uniqueness} and \emph{permanence}. These issues are addressed through subject-dependent EEG features which are recorded over temporally distinct recording days, 
	\item \emph{Reproducibility}. The recorded data are split into the training and validation data in two different setups, \emph{biased} and \emph{rigorous} setup,
	\item \emph{Fast response}. The classification is performed by both a fast non-parametric (cosine distance) and standard parametric approaches (linear discriminant analysis and support vector machine).
\end{enumerate}
Through these distinctive features, we successfully introduce a proof-of-concept for a wearable in-ear EEG biometrics in the community.

    \begin{figure}[b]
\begin{center}
	\includegraphics[clip, width=\linewidth]{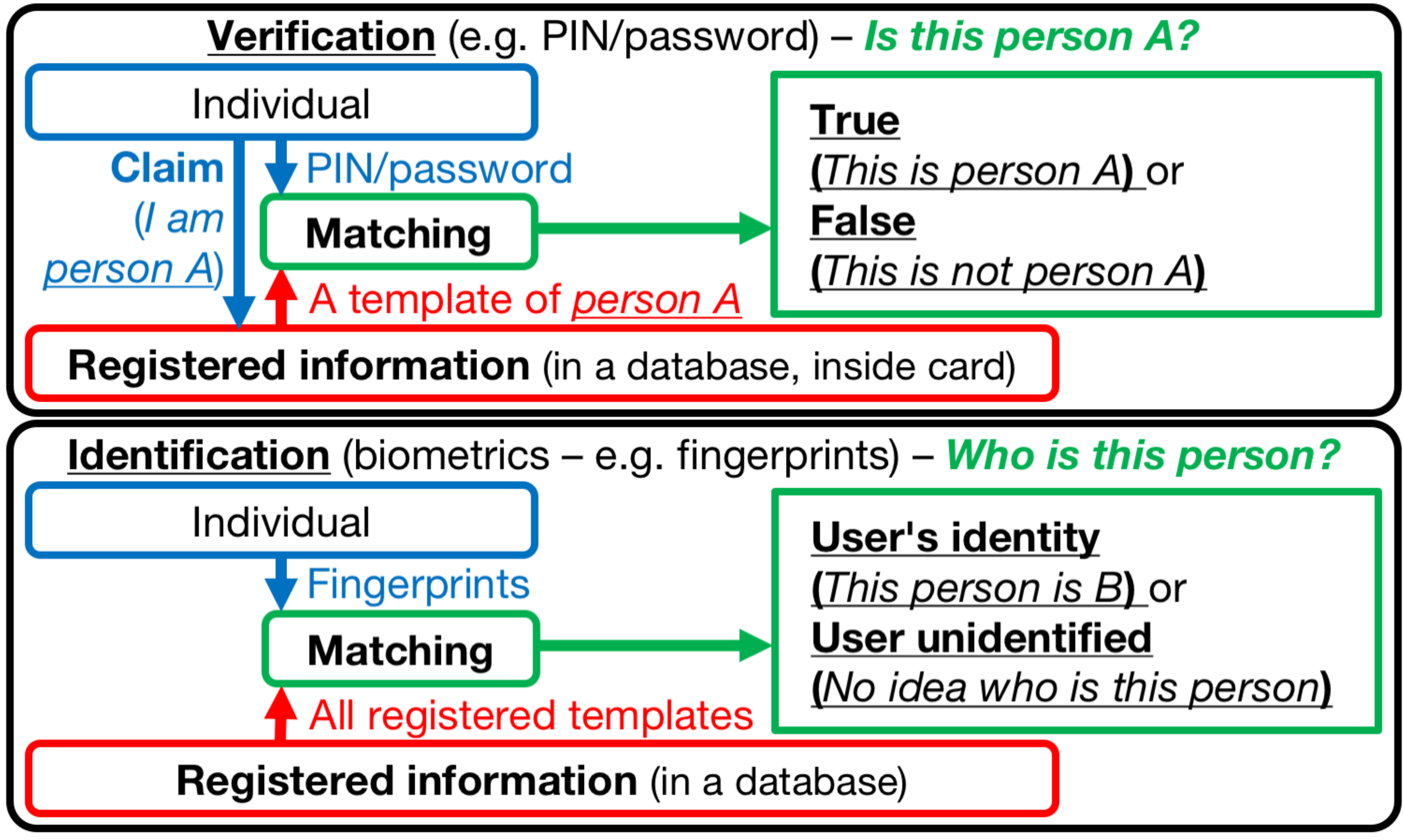}
    	\caption{{\color{black}Person recognition systems. \emph{Top}: Verification system. \emph{Bottom}: Identification system.}}
	\label{fig:bio_overvirew}
\end{center}
	\end{figure}


\section{Overview of EEG based biometrics}

\subsection{Biometric systems with verification/identification}
Depending on the context, the two categories of biometric systems are: 1) verification systems and 2) identification systems, summarised in Figure \ref{fig:bio_overvirew} \cite{Prabhakar2003}. Verification refers to validating a person's identity based on their individual characteristics, which are stored/registered on a server. In technical terms, this type of a biometric system performs a \emph{one-to-one matching} between the `claimed' and `registered' data, in order to determine whether the claim is true. In other words, the question asked for this application is `Is this person A?' as illustrated in Figure \ref{fig:bio_overvirew} (top panel). In contrast, an identification system confirms the identify of an individual from cross-pattern matching of the all available information, that is, based on \emph{one-to-many template matching}. The underlining question for this application is `Who is this person?' as illustrated in Figure \ref{fig:bio_overvirew} (bottom panel).

\subsection{Feasible EEG biometrics design} \label{sec:feasible_application}
Traditionally, EEG-based biometrics research has been undertaken based on both publicly available datasets \cite{Palaniappan2007} and custom recordings as part of research efforts \cite{Campisi2014}.
However, most of existing studies failed to rigorously address the key criterion, \emph{collectability}, which is also related to repeatability.
A large number of studies, especially those conducted at the dawn of EEG biometrics research, employed classification of the clients based on supervised learning with the training and validation data coming from the same recording trial. 
However, this experimental setup cannot truly evaluate the performance in identifying individual features, since such classification does not take into account the varying characteristic among multiple recording trials and recording days. 
In addition, EEG is prone to contamination by artefacts from subjects' movements (e.g. eye blinks, chewing), while the sources of external noise include electrode noise, power line noise, and electromagnetic interference from the surroundings.
This opens the possibility to additionally incorrectly associate `EEG patterns' with either trial-dependent features or so-called noise-related features -- in other words, this setup is \emph{biased} in favour of high classification rate.
Therefore, given the notorious variability of EEG patterns across days, biometrics studies based on a single recording day (even for a single subject) can only validate very limited scenarios, without any notion of repeatability and long-term feasibility \cite{Campisi2014}. 

\begin{figure}[tb]
\begin{center}
	\includegraphics[clip, width=\linewidth]{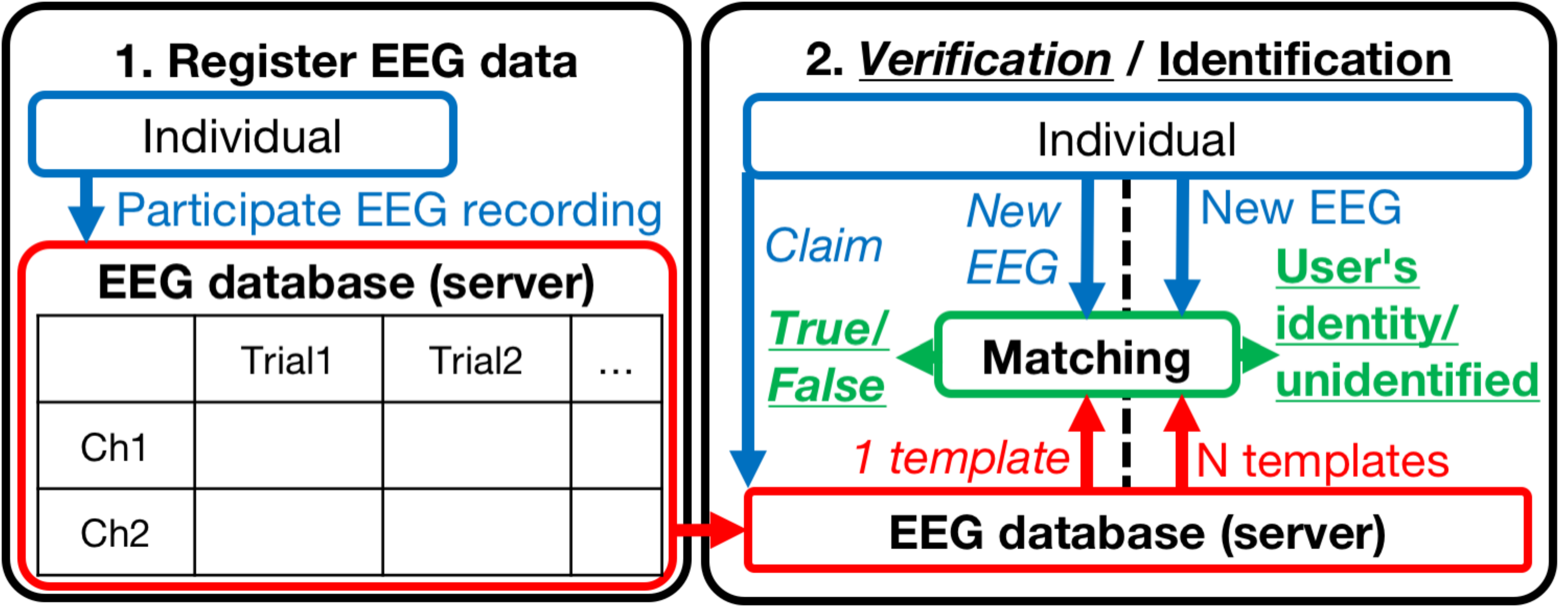}
    	\caption{Feasible EEG biometrics verification framework. \emph{Left}: EEG recording registration. The registered EEG signal must have been recorded beforehand. \emph{Right}: Verification and identification system.}
	\label{fig:bio_overvirew_1}
\end{center}
	\end{figure}

Figure \ref{fig:bio_overvirew_1} shows the concept of a \emph{rigorous} EEG biometrics verification/identification system in the real-world. Individuals participate in EEG recordings and their EEG signals are registered and stored on a server or in a database (left panel).
The client is granted access to their account by providing new EEG data in verification scenarios, whereby the algorithm discriminates the identify of an individual is in identification scenarios through new EEG recordings. Recall that the registered EEG must be recorded beforehand.

In order to fulfil the feasibility requirement, several studies performed successful EEG-based biometrics from multiple recording trials conducted on multiple distinct days, thus satisfying the \emph{collectability} requirement. 
However, the majority of these studies were still conducted in an unrealistic scenario, whereby the training and validation data in the classification process are split into segments, with all the segments coming from multiple trials but on the same recording day being randomly assigned to the training and validation datasets.
%
%
Therefore, this \emph{biased} setup, despite being based on the classification from multiple recording trials, mixes the training and test recordings from the same recording day and thus cannot truly evaluate the performance in the identification of individual features. 

In order to truly validate the robustness of a EEG biometrics application, within a \emph{rigorous} setup, it is therefore necessary to both: i) conduct multiple recordings over multiple days, and ii) to assign recordings on one day as the training data and use the recordings from the other days as the validation data. In other words, the training and validation datasets should be created \emph{so as not to share the same recording days} (as illustrated later in Figure \ref{fig:validation}, Setup-R). As emphasised by Rocca {\it et al.}, the issue of the repeatability of EEG biometrics in different recording sessions is still a critical open issue \cite{Rocca2014}. 

\begin{table*}[h]
\centering
\caption{Recording and classification set-up, and performance comparison among existing EEG biometrics}
\label{table:comparison}
\begin{tabular}{P{0.5cm} P{0.6cm}  P{0.7cm} P{0.5cm} P{2.4cm} c c P{0.7cm} c cc}
\hline
 Setup&Author &Subjects &Days  & Interval & Ch.  &Task$^\ast$&Segment &Feature Extraction&Classifier$^{\ast\ast}$&Performance \\ 
 \hline
  \multirow{4}{*}{\normalsize \rot{\emph{\textbf{{Biased}}}}} & \multirow{2}{*}{\cite{Abdullah2010}}& \multirow{2}{*}{10}  & \multirow{2}{*}{5} &  \multirow{2}{*}{\unit[2]{weeks}} &  \multirow{2}{*}{4}  &EC&  \multirow{2}{*}{\unit[5]{s}}& \multirow{2}{*}{AR}& \multirow{2}{*}{ANN}&CRR = \unit[97.0]{\%}\\
  	&		&  & &  & &EO& &&&CRR = \unit[96.0]{\%}\\
\cdashline{2-11}			
& \cite{Riera2008}&51&4& 34 $\pm$ \unit[74]{days}& 2 &EC & \unit[4]{s} & AR, PSD, MuI, COH, CC & FDA & EER = \unit[3.4]{\%}\\
 \cdashline{2-11}			
  &\cite{Su2010}&40&2& - & 1 &EC & \unit[180]{s} & AR, PSD & KNN, LDA & CRR = \unit[97.5]{\%}\\
  \hline
 \multirow{8}{*}{\normalsize \rot{\emph{\textbf{Rigorous}}}}& \cite{Marcel2007}&9&3&\unit[3]{days} & 8 &MI & \unit[1]{s} & PSD & MAP model & HTER = \unit[19.3]{\%}\\
 \cdashline{2-11}			
&\cite{Lee2013}&4&2& \unit[10]{days} - \unit[5]{months} & 1 & EC & \unit[50]{s}& PSD&  LDA & AC = \unit[100]{\%} \\
\cdashline{2-11}			
&  \multirow{2}{*}{\cite{Rocca2013}} & \multirow{2}{*}{9} &  \multirow{2}{*}{2} &  \multirow{2}{*}{1 - \unit[3]{weeks}} & 3 &  \multirow{2}{*}{EC} &  \multirow{2}{*}{\unit[1]{s}} &  \multirow{2}{*}{AR}  &  \multirow{2}{*}{Linear classier} &  CRR = \unit[100]{\%}\\
&			 &  &  &  & 5 &  &  & &  & CRR = \unit[100]{\%}\\
			 \cdashline{2-11}			
  &\multirow{2}{*}{\cite{Armstrong2015}}& 15 &  2 &  5 - \unit[40]{days} &  \multirow{2}{*}{1 }&  \multirow{2}{*}{ERP}&  \multirow{2}{*}{\unit[1.1]{s}} &  \multirow{2}{*}{Time-series}  &  \multirow{2}{*}{CC} & CRR = \unit[89.0]{\%}\\
&				  & 9 & 3 & 134 - \unit[188]{days} &  & &    &   & & CRR = \unit[93.0]{\%}\\
				  \cline{2-11}			
& \multirow{2}{*}{\cite{Maiorana2016a}}& \multirow{2}{*}{50}&  \multirow{2}{*}{3} &  \multirow{2}{*}{Ave. \unit[34]{days}}&\multirow{2}{*}{19} & EC&  \multirow{2}{*}{\unit[45]{s}} &  \multirow{2}{*}{AR, PSD, COH}&   \multirow{2}{*}{L1, L2, Cos. dist.}& R1IR = \unit[90.8]{\%}\\
& &  &   &   & & EO&   &   &   &  R1IR = \unit[85.6]{\%}\\	
 \hline
 \multicolumn{2}{c}{Task$^\ast$} &\multicolumn{9}{l}{EC: resting state with eyes closed, EO: resting state with eyes open, MI: motor imagery, ERP: event related potential}\\
 \multicolumn{2}{c}{\multirow{2}{*}{Classifier$^{\ast\ast}$}} &\multicolumn{9}{l}{ANN: artificial neural networks, FDA: Fisher discriminant analysis, KNN: k-nearest neighbours, LDA: linear discriminant analysis}\\
  & &\multicolumn{9}{l}{MAP: maximum a posteriori, CC: cross correlation, L1 (Manhattan) distance, L2 (Euclidean) distance, cosine distance}\\
 \hline
\end{tabular} 
\end{table*}


\subsection{Previous protocols} \label{sec:previous_protocol}
Table \ref{table:comparison} summarises the state-of-the-art of the existing EEG biometrics applications based on multiple data acquisition days. 

\emph{\textbf{Biased setup.}} In the first category (Setup: \emph{biased}) is the studies where the training and validation features were randomly selected regardless of the data acquisition days.
Abdullah {\it et al.} \cite{Abdullah2010} collected 4 channels of EEG data from 10 male subjects during the resting state, in the both eyes open (EO) and eyes closed (EC) scenarios, in 5 separate recording days over a course of 2 weeks. In each recording day, 5 trials of \unit[30]{s} recordings were recorded, and the recorded data were split into \unit[5]{s} segments with an overlap of \unit[50]{\%}. The autoregressive (AR) coefficients of the order $p = 21$ were extracted from each segment, and the extracted features were \emph{randomly} divided into the training (\unit[90]{\%}) and validation (\unit[10]{\%}) sets, namely 10-fold cross-validation. An artificial neural network (ANN) yielded \unit[97.0]{\%} of correct recognition rate (CRR) for the EC task, and \unit[96.0]{\%} of CRR for the EO task. 
Riera {\it et al.} \cite{Riera2008} recorded 2 forehead channels of EEG from 51 subjects over 4 separate recording days. The average temporal distance between the 1st and the 4th recording was 34 $\pm$ 74 days. The participants performed the EC task, and the duration of recordings was between 2 and 4 minutes. The recorded EEG was split into \unit[4]{s} segments, and five types of features were calculated for each segment, namely AR coefficients of order $p = 100$, power spectral density (PSD) in the \unit[$1-40$]{Hz} band, mutual information (MuI), coherence (COH), and cross-correlation (CC). The authors trained various classifiers and identified the 5 best classifiers. The Fisher's discriminant analysis (FDA) was then employed and was first trained with different types of discriminant functions using the 1st to 3rd recording trials; then the 4th recording trials were used for testing. Next, the best 5 classifiers from the training process were utilised for authentication tests, using the first and the second minutes of recordings from each trial; therefore, \emph{the training data for the classifiers and test data for the validation were not disjoint} (biased setup). The discriminant analysis with the selected discriminant function achieved \unit[3.4]{\%} of equal error rate (EER). 
Su {\it et al.} \cite{Su2010} analysed 5 minutes of the EC task from 40 subjects, with 6 recording trials performed in 2 separate recording days for each subject. The recorded EEG data from the FP1 channel were split into segments of multiple lengths. The PSD in the \unit[$5-32$]{Hz} band and AR coefficients of the order $p = 19$ were chosen as features. 
The extracted features were \emph{randomly} divided into the training (\unit[50]{\%}) and validation (\unit[50]{\%}) sets. As a result of 100 iterations, the classifier combining Fisher's linear discriminant analysis (LDA) and k-nearest neighbours (KNN) achieved an average CRR = \unit[97.5]{\%}, for a segment length of \unit[180]{s}.

\emph{\textbf{Rigorous setup.}} Multiple research groups considered EEG biometrics based on splitting the training and validation data in a \emph{rigorous} way, \emph{so as not to share the data from the same recording days} to highlight the feasibility of their system (Setup: \emph{rigorous}). 
Marcel {\it et al.} \cite{Marcel2007} analysed 8 channels of EEG from 9 subjects, with 4 recording trials over 3 consecutive days. The \unit[15]{s} trials consisted of two different motor imagery (MI) mental tasks, the imagination of hand movements. The recorded data were split into \unit[1]{s} segments, and PSD in the \unit[$8-30$]{Hz} band was calculated for each segment. The Gaussian mixture model (GMM) was chosen as a classifier, and maximum a posteriori (MAP) estimation was used for adapting a model for client data. By combining recordings over two days as training data, the authors achieved \unit[19.3]{\%} of half total error rate (HTER), which is a performance criterion widely used in biometrics; for more detail see Section \ref{sec:Performance_eval}. 
Lee {\it et al.} \cite{Lee2013} conducted an experiment of \unit[300]{s} in duration from four subjects over two days, based on single channel of EEG in the EC scenario. The data were segmented into multiple window sizes, and to extract frequency domain features, PSD was calculated only for the $\alpha$ band (\unit[$8-12$]{Hz}). Even though the dataset size was relatively small, with \unit[50]{s} of segment length, the LDA achieved \unit[100]{\%} classification accuracy. 
Rocca {\it et al.} \cite{Rocca2013} recorded two resting state EEGs, in both the eyes open (EO) and eyes closed (EC) scenarios, from 9 subjects over 2 different recording days, which were spanned 1 to 3 weeks. The recording length was \unit[60]{s}, and the recorded data were split into \unit[1]{s} segments with an overlap of \unit[50]{\%}. The AR model (Burg algorithm) of the order $p = 10$ was employed for feature extraction, and the training and validation data were split without mixing the trials from different recording days. The recognition results, with features from selected 3 or 5 channels of scalp EEG, were obtained by linear classification based on minimising the mean square error (MMSE), to achieve CRR = \unit[100]{\%}. 
Armstrong {\it et al.} \cite{Armstrong2015} recorded event related potentials (ERPs) and constructed two datasets. One dataset included EEG recorded from 15 subjects in two separate days, with a 5 - 40 inter-day interval, and the other one contained EEG from 9 subjects obtained in three separate days, with the average interval between the first and third recordings of 156 days. The recorded data from the O2 channel were split into \unit[1.1]{s} long segments, which contained an ERP and started from a \unit[100]{ms} pre-stimulus. The cross-correlation (CC) between the training and validation data was used as a feature for classification, and CRR = \unit[89.0]{\%} was achieved for validating the 2nd day recordings whereas CRR = \unit[93.0]{\%} for classifying the 3rd day recordings. 
Maiorana {\it et al.} \cite{Maiorana2016a} analysed 19 channels of EEG from 50 subjects during both EC and EO tasks in three different recording days, with the average interval between the first and the third recording of 34 days. Each recording trial consisted of \unit[240]{s} of data, segmented into \unit[5]{s} windows, with an overlap of \unit[40]{\%}. Three types of features were extracted, including a channel-wise AR model (using the Burg algorithm) of the order $p = 12$, channel-wise PSD, and the coherence (COH) between the EEG channels. The L1, L2, and cosine distances were calculated for the extracted features, and the rank-1 identification rate (R1IR) achieved \unit[90.8]{\%} accuracy in the EC task and \unit[85.6]{\%} in the EO task.
%


\subsection{Biometrics based on collectable EEG systems}
With a perspective of \emph{collectability}, a biometrics application with dry EEG electrodes was recently introduced \cite{Chen2016}. While conventional wet EEG headsets require the application of a conductive gel which is generally time-consuming, the dry headset with 16 scalp channels took on the average 2 minutes to be operational. The brain-computer interface based biometrics application with rapid serial visual presentation paradigm achieved CRR = \unit[100]{\%} with \unit[27]{s} window size over all 29 subjects. Although the recordings were performed over a single recording day per subject, the application with a dry headset was a step forward towards establishing collectable EEG biometrics in real-world.

In a recent effort to enable collectable EEG, the in-ear sensing technology \cite{Looney2012} was introduced into the research community. The ear-EEG has been proven to provide on-par signal quality, compared to conventional scalp-EEG, in terms of steady state responses \cite{Looney2012, Kidmose2013a}, monitoring sleep stages \cite{Looney2016a, Nakamura2017a}, and also for monitoring cardiac activity \cite{Goverdovsky2017, Goverdovsky2015}. The advantages of the in-ear EEG sensing for a potential biometrics application in the real-world are: 
\begin{itemize}
\item \emph{Unobtrusiveness}: The latest `off-the-shelf' generic viscoelastic EEG sensor is made from affordable/consumable standard earplugs \cite{Goverdovsky2016},
\item \emph{Robustness}:  The viscoelastic substrate expands after the insertion, so the electrodes fit firmly inside the ear canal \cite{Goverdovsky2017}, {\color{black} where the position of electrodes remains the same in different recording sessions,}
\item \emph{User-friendliness}: The sensor can be applied straightforwardly by the user, without the need for a trained person.
\end{itemize}
Therefore, biometrics with ear-EEG offers a high degree of \emph{collectability}, a critical issue in real-world applications. {\color{black} Previously, even based on this biased scenarios, in-ear EEG based biometrics application has been proposed in \cite{Curran2016}.}

\subsection{Problem formulation}
We investigate the possibility of biometrics verification with a wearable in-ear sensor, which is capable of fulfilling the \emph{collectability} requirement. 
The data were recorded over temporally distinct recording days, in order to additionally highlight the \emph{uniqueness} and \emph{permanence} aspects. 
Although the changes in EEG rhythms may well depend on the time period of years rather than days, the alpha band features during the resting state with eyes closed were reported as the most stable EEG feature over two years \cite{Neuper2005}. Since EEG alpha rhythms predominantly originate during wakeful relaxation with eyes closed, we chose our recording task to be the resting state with eyes closed. This task was used in multiple previous studies \cite{Abdullah2010, Riera2008, Su2010,Lee2013,Rocca2013,Maiorana2016a}. 
In order to design a feasible biometrics application in the real-world, we considered imposters in two different ways: i) registered subjects in a database, and ii) subjects not belonging to a database. Previously, Riera {\it et al.} \cite{Riera2008} also used a single trial of EEG recording from multiple subjects as `intruders', while the `imposters' data were EEG recordings available from multiple other experiments. For rigour, we collected two types of data: 1) based on multiple recordings from {\color{black} fifteen} subjects over two days, and 2) multiple recordings from {\color{black} five} subjects, which were only used for imposters' data. 
The classification was performed by both a non-parametric and parametric approach. The non-parametric classifier, minimum cosine distance, is a simplest way for evaluating the similarity between the training and validation matrix, whereas the parametric approach, the support vector machine (SVM), was tuned within the training matrix in order to find optimal hyper-parameters and weights for validation. The same hyper-parameters and weights were used for classifying the validation matrix. Besides, the linear discriminant analysis (LDA) was also employed as a classifier. Through the binary client-imposter classification, we then evaluated the feasibility of our in-ear EEG biometrics.

\section{Methods}

\subsection{Data acquisition}
The recordings were conducted at Imperial College London, for two different groups of subjects and under the ethics approval, Joint Research Office at Imperial College London ICREC12\_1\_1.  One set of data were the recordings used as both clients and imposters data, denoted by $S_R$, and the other subset were the recordings for only imposters' data, denoted by $S_N$. Table \ref{table:recording_details} summarises the two recording configurations of $S_R$ and $S_N$. 

\begin{figure}[bp]
\begin{center}
    		\includegraphics[clip, width=\linewidth]{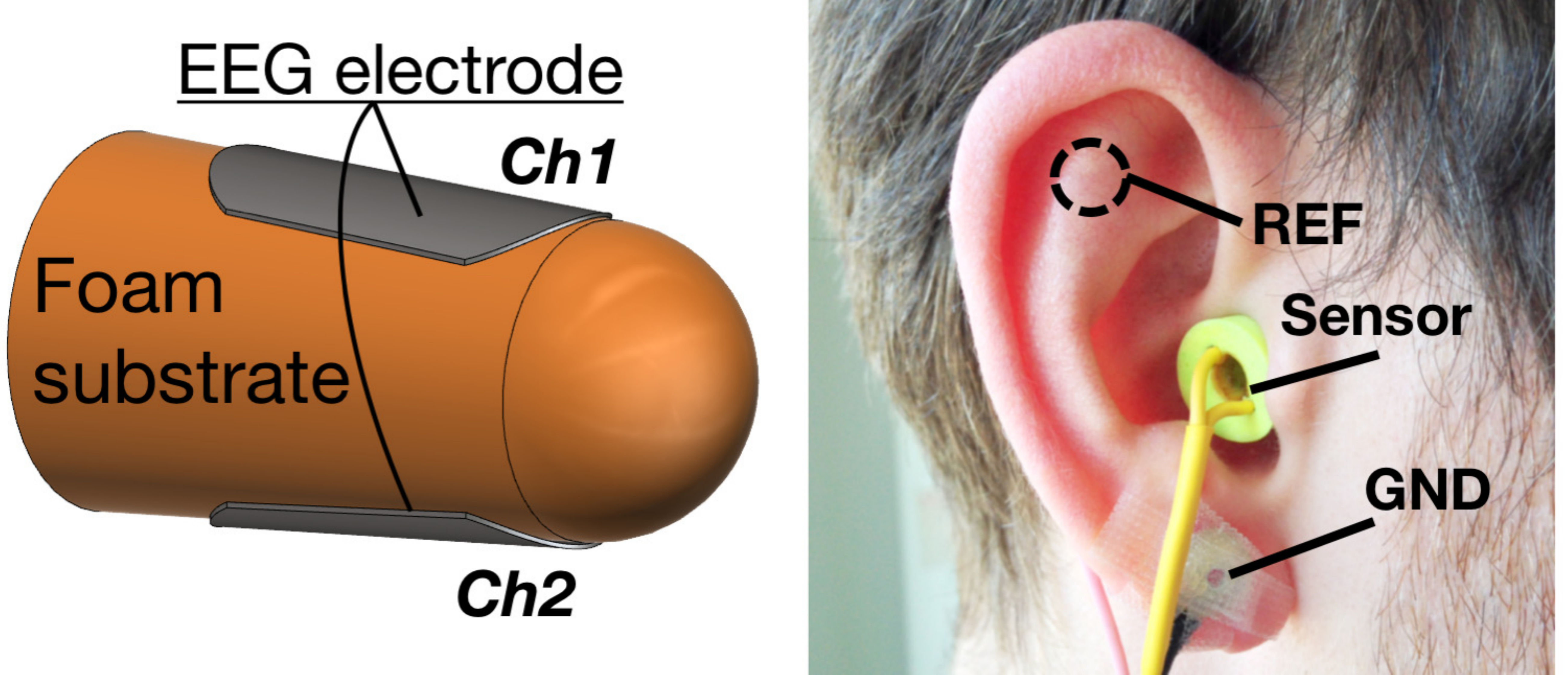}
    	\caption{The in-ear sensor used in our study. \emph{Left}: Wearable in-ear sensor with two flexible electrodes. \emph{Right}: Placement of the generic viscoelastic earpiece.}
	\label{fig:earpiece}
\end{center}
	\end{figure}

For the $S_R$ subset of recordings, {\color{black} fifteen} healthy male subjects (aged {\color{black}22}-38 years) participated in two temporally separate sessions, with the interval between two recording sessions between 5 and {\color{black}15} days, depending on the subject.
The participants were seated in a comfortable chair during the experiment, and were asked to rest with eyes closed. The length of each recording was \unit[190]{s}, and the recording was undertaken three times (trials) per one day. 
The interval between each recording trial was approximately 5 to 10 minutes. 
In total, six trials were recorded per subject. 
The in-ear sensor was inserted in the subject's left ear canal after earwax was removed; it then expanded to conform to the shape of the ear canal. The reference gold-cup standard electrodes were attached behind the ipsilateral earlobe and the ground electrodes were placed on the ipsilateral helix. For simplicity, the upper electrode is denoted by Ch1, while Ch2 refers to the bottom electrode, as shown in Figure \ref{fig:earpiece} (left panel).
The two EEG signals from flexible electrodes were recorded using the g.tec g.USBamp amplifier with a 24-bit resolution, at a sampling frequency $fs$ = \unit[1200]{Hz}.

For the $S_N$ subset of recordings, {\color{black} five} healthy subjects (aged {\color{black}22-29} years) participated in three recording trials. Similar to the $S_R$ subset of recordings, the participants were seated in a comfortable chair, and were resting with eyes closed. The duration of recording was also {\color{black} \unit[190]{s}}. A generic earpiece with two flexible electrodes \cite{Goverdovsky2016} was inserted in the subject's {\color{black} left} ear canal and the same reference and ground configuration was utilised for the $S_R$ subset of recordings. 

{\color{black} Similar to the setup in \cite{Maiorana2016a}, there was no restriction on the activities that the subjects performed, and no health test such as their diet and sleep, was carried out neither before or between an EEG acquisition and the following one, nor during the days of the recordings. This lack of restrictions allowed us to acquire data in conditions close to real life.}

\begin{table}[bp]
\centering
\caption{Two EEG recordings and corresponding subset}
\label{table:recording_details}
\begin{tabular}{ccc}
\hline
  Subset       & Client/Imposter $S_R$  & Imposter only $S_N$\\
         \hline
  Task&     \multicolumn{2}{c}{Resting state with eyes closed}\\
  No. Subjects   & {\color{black} 15} & {\color{black} 5}\\
No. Trials& 6 (2 days, 3 trials) & {\color{black} 3 (1 day, 3 trials)}\\
Duration &  \multicolumn{2}{c}{\color{black} \unit[190]{s}}\\
              \hline
              \end{tabular}
\end{table}

    \begin{figure}[htbp]
\begin{center}
    		\includegraphics[clip, width=\linewidth]{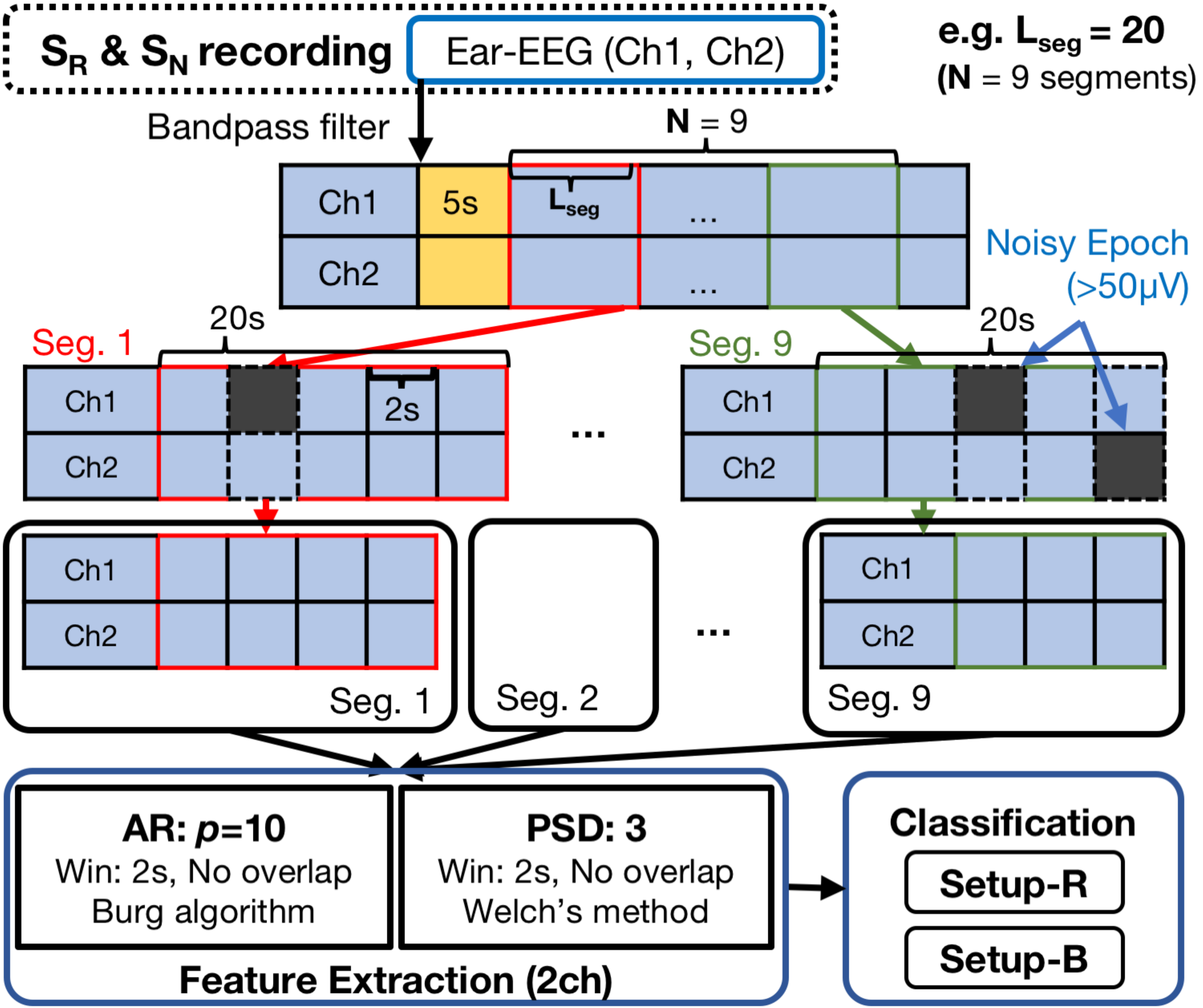}
    	\caption{\color{black}Flowchart for the biometrics analysis framework in this study.}
	\label{fig:protocol}
\end{center}
	\end{figure}

{\color{black}
\subsection{Ear-EEG sensor}
 The in-ear EEG sensor is made of a memory-foam substrate and two conductive flexible electrodes, as shown in Figure \ref{fig:earpiece}. The substrate material is a viscoelastic foam, therefore the `one-fits-all' generic earpiece fits any ear regardless of the shape. The size of earpiece was the same for over twenty subjects (both the $S_R$ and $S_N$ subjects). Further details of the construction of such a viscoelastic earpiece and its detailed recordings of various brain functions can be found in \cite{Goverdovsky2016, Goverdovsky2017}.}

\subsection{Pre-processing}
The two channels of the so-obtained ear-EEG were analysed based on the framework illustrated in Figure \ref{fig:protocol}. In each recording, for both the $S_R$ and $S_N$ recordings, the first \unit[5]{s} of recording data were removed from the analyses, in order to omit {\color{black} noisy recordings arising at the beginning of the acquisition}. The two recorded channels of EEG were bandpass filtered with the fourth-order Butterworth filter with the pass-band \unit[$0.5-30$]{Hz}. The bandpass filtered signals were split into segments. {\color{black} The symbol $N$ denotes the number of segments per recording trial from both the $S_R$ and $S_N$ subsets. The lengths of segments were chosen as $L_{seg} = 10, 20, 30, 60, 90$ s. Therefore, when the segment length was $L_{seg} = $ \unit[60]{s}, $N=3$ ($190-5 = 185$ s, $\floor*{185/60} =  3$) segments were extracted from every recording trial of the $S_R$ and $S_N$. Within each segment, the data was split into epochs of \unit[2]{s} length.
The epochs with the amplitudes of greater than \unit[50]{$\mu$V} for either Ch1 or Ch2 were considered corrupted by artefacts and removed from the analyses. This method resulted in a loss of 4.3 \% of the data, namely approximately \unit[7.7]{s} out of \unit[190]{s} per recording trial.}

\subsection{Feature Extraction}
After the pre-processing, two types of features were extracted from each segment of the ear-EEG. For a fair comparison with the state-of-the-art, these features were selected to be the same or similar to those used in the recent studies based on the resting state with eyes closed \cite{Lee2013, Rocca2013}, and included: 1) a frequency domain feature -- power spectral density (PSD), and 2) coefficients of an autoregressive (AR) model. 

\subsubsection{The PSD features}
\begin{figure}[tb]
\begin{center}
		\includegraphics[clip, width=\linewidth]{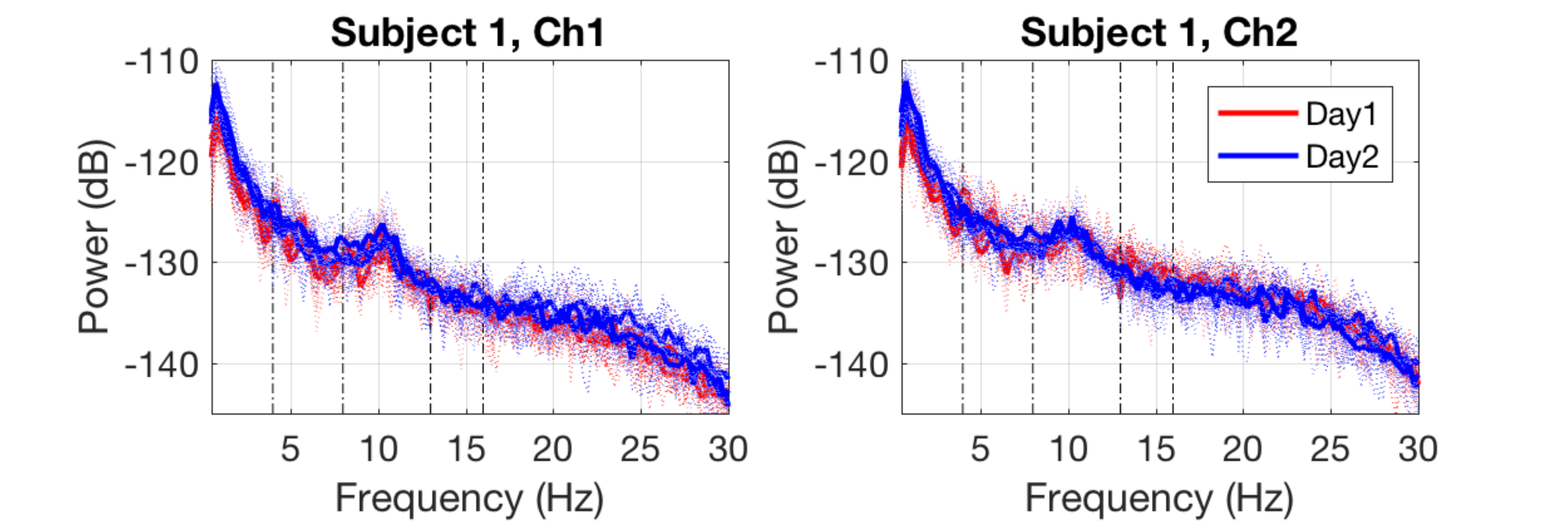}
		\includegraphics[clip, width=\linewidth]{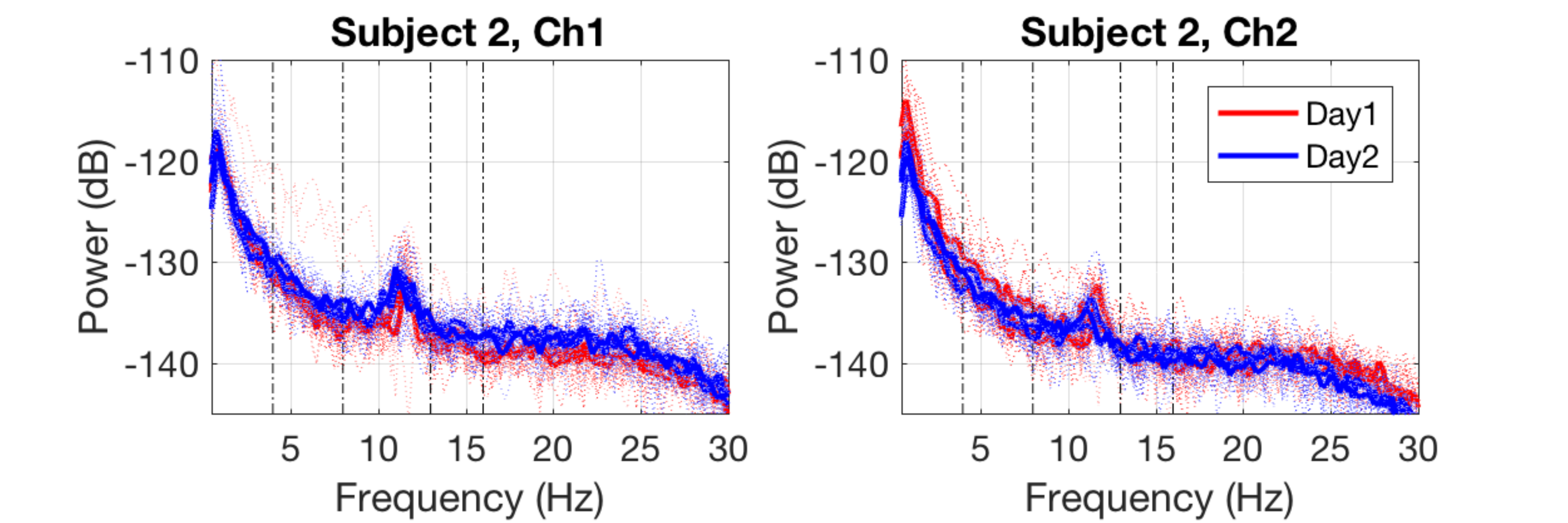}
    	\caption{{\color{black}Power spectral density for the in-ear EEG Ch1 (left) and the in-ear EEG Ch2 (right) of Subject 1 (top panels) and Subject 2 (bottom panels). The thick lines correspond to the averaged periodogram obtained by the all recordings from the 1st day (red) and the 2nd day (blue), whereas the thin lines are the averaged periodogram obtained by a single trial.}}
	\label{fig:psd_sample}
\end{center}
	\end{figure}

{\color{black} Figure \ref{fig:psd_sample} shows power spectral density for the in-ear EEG Ch1 (left) and for the in-ear EEG Ch2 (right) of two subjects. For this analysis, the recorded signals were conditioned with the fourth-order Butterworth filter with the pass-band \unit[$0.5-30$]{Hz}. The PSD were obtained using Welch's averaged periodogram method \cite{hayes1996statistical}, the window length was \unit[20]{s} with \unit[50]{\%} of overlap. The PSDs are overlaid between different recording days (red: Day1, blue: Day2), as well as among different recording trials with the same recording days, especially visible from 3 to \unit[$20$]{Hz}. 
Previously, Maiorana {\it et al.} utilised PSD features for EEG biometrics based on the resting state eyes closed and achieved the best performance between the PSD features from theta to beta band, which was classified by the minimum cosine approach \cite{Maiorana2016a}; the inclusion of the delta band decreased their identification performance. In our in-ear EEG biometrics approach, the obtained PSDs were visually examined and we found that the ratio between the the total $\alpha$ band (\unit[$8-13$]{Hz}) power and the total $\theta-\alpha_{high}$ band (\unit[$4-16$]{Hz}) power is a relatively more significant individual factor for biometrics, rather than the total $\alpha$ band (\unit[$8-13$]{Hz}) power, which is proposed in \cite{Lee2013}. Therefore,} in each segment of length $L_{seg}$, univariate PSD was calculated by Welch's method with \unit[2]{s} of the window length and no overlap. Three features were obtained for each PSD: 
 1) The ratio between the total $\alpha$ band (\unit[$8-13$]{Hz}) power and the total $\theta-\alpha_{high}$ band (\unit[$4-16$]{Hz}) power, 
 2) the maximum power in $\alpha$ band, and 
 3) the frequency corresponding to the maximum of $\alpha$ band power. In total, $D = 6$ (three features $\times$ two channels) frequency domain features were extracted from each segment. 


\subsubsection{The AR features}
The Burg algorithm \cite{hayes1996statistical} of order $p = 10$ was used to estimate the AR coefficients. For each segment, we applied univariate AR parameter estimation of its $\alpha$ band (\unit[$8-13$]{Hz}) with a window length of \unit[2]{s} and no overlap. The AR model was chosen as a feature, because it was used in a previous successful study on EEG biometrics based on the resting state with eyes closed \cite{Rocca2013}. A total of $D = 20$ features (ten coefficients $\times$ two channels) were therefore extracted for each ear-EEG segment.

\subsection{Validation scenarios}

{\color{black} With the extraction of the both univariate AR and PSD features from two channels, the dimension $D$ of features per EEG segment was twenty six. Recall that the first \unit[5]{s} of recording data were removed from the analyses. For each trial of the $S_R$ and $S_N$ recordings, the data with the duration of \unit[190]{s} was split into segments of length $L_{seg} = 10, 20, 30, 60, 90$ s. Therefore, $N = 18, 9, 6, 3, 2$ segments were respectively obtained. Each recording trial was represented by the feature matrix $X_R$ for the $S_R$ recordings and $X_N$ for the $S_N$ recordings, such matrices have $N \times D$ elements. 
In this way, a set of six feature matrices was obtained from one subject for the $S_R$ recordings (three recording trials per one day, over two different days), whilst a set of three feature matrices was obtained from one subject for the $S_N$ recordings (three recording trials per one day, one recording day).}
We next discuss the use of feature matrices in two different validation scenarios.

As emphasised in Introduction, we introduce a feasible EEG biometrics which satisfies the \emph{collectability} requirements, which are also related to repeatability.
Therefore, for rigour, we used all feature matrices $X_R$ from the 1st day of recordings as the training data, and feature matrices from the 2nd day of recordings as the validation data, and vice versa (Setup-R)\footnotemark. Our goal was to examine the robustness of the proposed approach over the two different time periods in Setup-R.
For the second setup, Setup-B\footnotemark[\value{footnote}], training feature matrices were also selected from the trials which were recorded at the same recording day as the validation matrices. Namely, the training and validation data are split by mixing the data from the same recording days. {\it Notice that, although used in most available EEG biometrics studies \cite{Abdullah2010,Riera2008,Su2010}, Setup-B could not evaluate the repeatability/reproducibility of the application, because the training and validation data were both from the same recording days.} In other words, such an approach benefits from the recording-day-dependent EEG characteristic in the classification. However, as the number of feasible biometric modalities with in-ear EEG sensor is limited and for comparison with other studies, for convenience we also provide the results for Setup-B.
\footnotetext{Setup-\emph{R}: \emph{Rigorous} setup, Setup-\emph{B}: \emph{Biased} setup.}

\begin{algorithm*}
\caption{, {\bf Setup-R (rigorous):} Select the training and validation data {\bf without mixing} segments from the two recording days, e.g. $[i,j,k] = $ [1,1,1]}
\label{setup-A}
\begin{algorithmic}[1]
\State \begin{varwidth}[t]{\linewidth}
	$VC$: The matrix of the selected-subject, the selected-day and the selected-trial, [1,1,1] \par
	\hspace{1cm} $Y_{VC} = X_R^{(1, 1, 1)}$
\end{varwidth}
\vspace{0.1cm}
\State \begin{varwidth}[t]{\linewidth}
	$VI$: The matrices of the selected-trial and the selected-day from the non-selected-subjects, [(2:15),1,1] \par
	\hspace{1cm} $Y_{VI} = [X_R^{(2,1,1)T}, X_R^{(3,1,1)T}, ..., X_R^{(15,1,1)T}]^T$
\end{varwidth}	
\vspace{0.1cm}
\State \begin{varwidth}[t]{\linewidth}
	$TC$: The matrices of all recording trials recorded at the non-selected-day from the selected-subject, [1,2,(1:3)] \par
	\hspace{1cm} $Y_{TC} = [X_R^{(1,2,1)T}, X_R^{(1,2,2)T}, X_R^{(1,2,3)T}]^T$
\end{varwidth}	
\vspace{0.05cm}
\State \begin{varwidth}[t]{\linewidth}
	$TI$: The matrices of all recording data recorded at the non-selected-day from the non-selected-subjects, [(2:15),2,(1:3)] \par
	\hspace{1cm} $Y_{TI} = [X_R^{(2,2,1)T},..., X_R^{(2,2,3)T}, X_R^{(3,2,1)T},...,X_R^{(15,2,3)T}]^T.$
\end{varwidth}	
\vspace{0.05cm}
\State \begin{varwidth}[t]{\linewidth}
	$VI_N$: {\color{black}The matrices of the selected-trial from the $S_N$ recording subjects, [(1:5),-,1] \par
	\hspace{1cm} $Y_{VI_{N}} = [X_N^{(1,-,1)T},X_N^{(2,-,1)T}, ... , X_N^{(5,-,1)T}]^T.$}
\end{varwidth}
\end{algorithmic}
\end{algorithm*}
\begin{algorithm*}
\caption{, {\bf Setup-B (biased):} Select the training and validation data {\bf with mixing} segments from the two recording days, e.g. $[i,j,k] = $ [2,2,2]}
\label{setup-B}
\begin{algorithmic}[1]
\State \begin{varwidth}[t]{\linewidth}
	$VC$: The matrix of the selected-subject, the selected-day and the selected-trial, [2,2,2] \par
	\hspace{1cm} $Y_{VC} = X_R^{(2,2,2)}$
\end{varwidth}
\vspace{0.1cm}
\State \begin{varwidth}[t]{\linewidth}
	$VI$: The matrices of the selected-trial and the selected-day from the non-selected-subjects, [(1,3:15),2,2] \par
	\hspace{1cm} $Y_{VI} = [X_R^{(1,2,2)T}, X_R^{(3,2,2)T}, ...,  X_R^{(15,2,2)T}]^T.$
\end{varwidth}
\vspace{0.1cm}
\State \begin{varwidth}[t]{\linewidth}
	$TC$: The matrices of all recording trials recorded at the non-selected day from the selected-subject, [2,1,(1:3)], and the matrices of non-selected trials recorded at the selected-day from the selected-subject, [2,2,(1,3)] \par
	\hspace{1cm} $Y_{TC} = [X_R^{(2,1,1)T}, X_R^{(2,1,2)T}, X_R^{(2,1,3)T}, X_R^{(2,2,1)T}, X_R^{(2,2,3)T}]^T.$
\end{varwidth}
\vspace{0.05cm}
\State \begin{varwidth}[t]{\linewidth}
	$TI$: The matrices of all recording data recorded at the non-selected-day from the non-selected-subjects, [(1,3:15),1,(1:3)], \par
	and the matrices of non-selected trials recorded at the selected-day from the non-selected-subjects, [(1,3:15),2,(1,3)] \par
	\hspace{1cm} $Y_{TI} = [X_R^{(1,1,1)T},X_R^{(1,1,2)T},X_R^{(1,1,3)T},X_R^{(1,2,1)T},X_R^{(1,2,3)T}, X_R^{(3,1,1)T},...,X_R^{(15,2,3)T}]^T.$
\end{varwidth}
\vspace{0.05cm}
\State \begin{varwidth}[t]{\linewidth}
	$VI_N$: {\color{black} The matrices of the selected-trial from the $S_N$ recording subjects, [(1:5),-,2] \par
	\hspace{1cm} $Y_{VI_{N}} = [X_N^{(1,-,2)T},X_N^{(2,-,2)T}, ... , X_N^{(5,-,2)T}]^T.$}
\end{varwidth}
\end{algorithmic}
\end{algorithm*}

\begin{figure}[htbp]
\begin{center}
    		\includegraphics[clip, width=\linewidth]{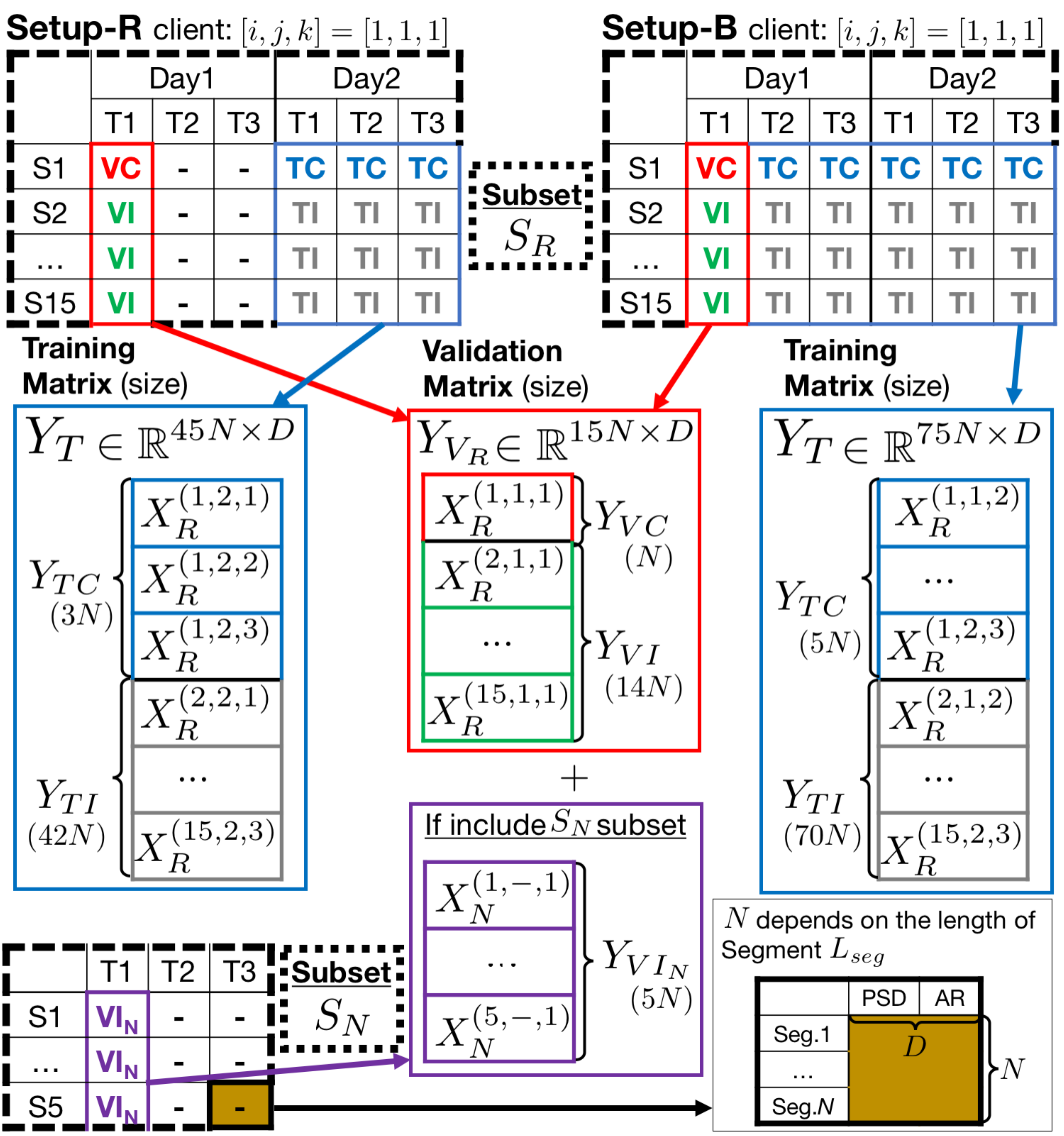}
    	\caption{{\color{black}Two validation scenarios (Setup-R and Setup-B), where $X_R^{(i,j,k)} \in \mathbb{R} ^ {N \times D}$ and  $X_N^{(i,-,k)} \in \mathbb{R} ^ {N \times D}$ denote a feature matrix from a single trial of the subject $i$, recording day $j$, and trial $k$, from the $S_R$ recordings and the $S_N$ recordings, respectively. The number of segments per recording trial $N$ depends on the chosen segment lengths $L_{seg}$. The dimension $D$ is the number of features per EEG segment.}}
	\label{fig:validation}
\end{center}
	\end{figure}

Figure \ref{fig:validation} summarises the two validation scenarios, Setup-R and Setup-B. For clarity, we denote by
$VC$ the validation feature matrix for the client,
by $VI$ the validation feature matrix for the imposters, 
while $TC$ is the training feature matrix for the client, 
and $TI$ as the training feature matrix for the imposters. 
The feature matrix from a single trial of the subject $i$, recording day $j$, and trial $k$, from the $S_R$ recordings is denoted by $X_R^{(i,j,k)}$.
Then, the training feature matrix $Y_{T}$ and the validation matrix $Y_{V}$ are given as
\begin{eqnarray*}
Y_{T} &=& [Y_{TC}^T, Y_{TI}^T] ^ T, \\
Y_{V} &=& Y_{V_{R}} = [Y_{VC}^T, Y_{VI}^T] ^ T.
\end{eqnarray*}
{\color{black} Besides, in order to evaluate feasibility in the real-world, we used $S_N$ recordings, which are EEG recordings only used for imposters; Riera {\it et al.} termed the imposter only data as `intruders' \cite{Riera2008}. For an additional scenario in both Setup-R and Setup-B (see Section \ref{sec:results4}), $S_N$ recordings were used as the validation data for imposters, $VI_N$. The feature matrix from a single trial of the subject $i$, and trial $k$, from the $S_N$ recordings is denoted by $X_N^{(i,-,k)}$. Therefore, the validation matrix is given by
\begin{eqnarray*}
Y_{V} &=& Y_{V_{R}} +  Y_{VI_{N}} =   [Y_{VC}^T, Y_{VI}^T, Y^T_{VI_{N}} ] ^ T.
\end{eqnarray*}}
Table \ref{table:data_structure} summarises the properties of matrices for the training matrix and the validation matrix in the both Setup-R and Setup-B. 

\begin{table}[htbp]
\centering
\caption{{\color{black} Dimensions of the training and validation matrix in Setup-R and Setup-B}}
\label{table:data_structure}
\begin{tabular}{P{0.4cm}P{0.3cm}cc}
\hline
&&Setup-R&Setup-B\\
\hline
\multirow{3}{*}{\rot{Train}}
&$Y_{TC}$&$3N \times D$&$5N \times D$\\
&$Y_{TI}$&$42N \times D$&$70N \times D$\\[0.2mm]
\cdashline{2-4}
&$Y_{T}$&$45N \times D$&$75N \times D$\\
\hline
\multirow{4}{*}{\rot{Validation}}
&$Y_{VC}$& \multicolumn{2}{c}{$N \times D$}\\
&$Y_{VI}$& \multicolumn{2}{c}{ $14N \times D$}\\
\cdashline{2-4}
&$Y_{V_{R}}$&\multicolumn{2}{c}{$15N \times D$}\\
\cline{2-4}
&$Y_{VI_{N}}$&\multicolumn{2}{c}{$5N \times D$ }\\[0.2mm]
              \hline \hline
\multicolumn{2}{c}{Total validation}&\multicolumn{2}{c}{$\begin{array} {lcl} Y_{V} &=& Y_{V_{R}}: 90(15N) \\Y_{V} &=& Y_{V_{R}} +  Y_{VI_{N}} : 90(15N + 5N) \end{array}$}\\
\multicolumn{2}{c}{elements}&\multicolumn{2}{c}{$i=$ 1:15 (Sub), $j =$ 1:2 (Day), $k =$ 1:3 (Trial)}\\              
              \hline
              \end{tabular}
\end{table}

\subsection{Classification} \label{sec:classification}
For both the Setup-R and Setup-B, we selected every trial from every subject for the validation of client data, so as to have validated ninety times (three trials $\times$ two days $\times$ fifteen subjects). For each validation, both the largest and smallest values were found for each feature (column-wise) from the training matrix, then the validation matrix was normalised to the range $[0, 1]$ based on these largest and smallest values. Three classification algorithms were employed: 1) a non-parametric approach -- minimum cosine distance \cite{Maiorana2016a}, 2, 3) parametric approaches -- {\color{black} linear discriminant analysis (LDA) \cite{murphy2012machine}} and support vector machine (SVM) \cite{Chang2011}.

\subsubsection{Cosine distance}
The cosine distance is the simplest way for evaluating the similarity between the rows of the validation matrix, {\color{black} $Y_{V_{(l,:)}}$, where $l = 1, ..., 15N$ for $Y_{V} = Y_{V_{R}}$ and $l = 1, ..., 20N$ for $Y_{V} = Y_{V_{R}} +  Y_{VI_{N}}$, and the training matrix, $Y_{T}$, }and is given by
\begin{eqnarray*}
d\left(Y_{V_{(l,:)}}, Y_{T}\right) = \min_{n} \frac{\sum_{m=1}^{D} Y_{V_{(l,m)}} Y_{T_{(n,m)}} }{\sqrt{\sum_{m=1}^{D} (Y_{V_{(l,m)}})^2}\sqrt{\sum_{m=1}^{D} (Y_{T_{(n,m)}})^2}}. 
\end{eqnarray*}
In other words, the cosine distance is used for evaluating the similarity between a given test sample (e.g $l$th row of the validation matrix, $Y_{V_{(l,:)}}$) and a template (training) feature matrix, $Y_{T}$. The distances between the $l$th row of the validation matrix, $Y_{V_{(l,:)}}$, and the each row of training matrix $Y_{T}$ were first computed, then the minimum among the computed distances was selected.  


\subsubsection{LDA}
{\color{black}The binary-class LDA was employed as a classifier. The LDA finds a linear combination of parameters to separate given classes. The LDA projects the data onto a new space, and discriminates between two classes by maximising the between-class variance while minimising the within-class variance.}

\subsubsection{SVM}
The binary-class SVM was employed as a parametric classifier \cite{Chang2011}. For both Setup-R and Setup-B, four hyper-parameters: type of kernel, regularisation constant for loss function $C$, inverse of bandwidth $\gamma$ of kernel function, and order of polynomial $d$, were tuned by 5-fold cross-validation within the training matrix. Then, the same hyper-parameters were used in order to obtain the optimal weight parameters within the training matrix. The same hyper-parameters and weight parameters as in the training were used for validation. Table \ref{table:kernels} summarises the hyper-parameters for SVM. 

\begin{table}[htbp]
\centering
\caption{Hyper-parameters for SVM}
\label{table:kernels}
\begin{tabular}{ccc}
\hline
  Type of kernel            &$\kappa(\mathbf{x},\mathbf{x}')$&       Hyper-parameters \\
 \hline
Linear    &  $ \mathbf{x}^T\mathbf{x}' $ & - \\
Sigmoid &  $\tanh(\gamma\mathbf{x}^T\mathbf{x}' + r) $  &  $\gamma, (r = 0)$ \\
RBF       & $\exp(-\gamma |\mathbf{x} - \mathbf{x}'|^2) $&  $\gamma$\\
Polynomial & $(\gamma \mathbf{x}^T \mathbf{x} + r)^d $&  $\gamma, d, (r = 0)$\\
              \hline
              \end{tabular}
\end{table}

\subsection{Performance evaluation} \label{sec:Performance_eval}
Feature extraction and classification with minimum cosine distance and with LDA was performed using Matlab 2016b, and the classification with SVM was conducted in Python 2.7.12 Anaconda 4.2.0 (x86\_64) operated on an iMac with 2.8GHz Intel Core i5, 16GB of RAM.

{\color{black} For the verification setup (the number of classes $M=2$, client-imposter classification), the performance was evaluated through the false accept rate (FAR), false reject rate (FRR), half total error rate (HTER), accuracy (AC), and true positive rate (TPR), defined as:
\begin{eqnarray*}
FAR = FP/(FP+TN), \,\, FRR = FN/(TP+FN), \,\, \,\, \,\, \,\, \\
 HTER = \frac{FAR + FRR}{2}, \,\, AC = \frac{TP+TN}{TP+FN+FP+TN}, \\
TPR = TP/(TP+FN). \,\, \,\, \,\, \,\,\,\, \,\, \,\, \,\, \,\,\,\, \,\, \,\, \,\,\,\, \,\, \,\, \,\,\,\, \,\, \,\, \,\, \,\,\,\, \,\, 
\end{eqnarray*}
The parameter TP (true positive) represents the number of positive (target) segments correctly predicted, TN (true negative) is the number of negative (non-target) segments correctly predicted, FP (false positive) is the number of negative segments incorrectly predicted as the positive class, and FN (false negative) is the number of positive segments incorrectly predicted as negative class. 

For the identification setup ($M=15$), the performance was evaluated by subject-wise sensitivity (SE), identification rate (IR) and Cohen's kappa ($\kappa$) coefficient as:
\begin{eqnarray*}
SE_{i} = TP_i/(TP_i+FN_i), \,\, IR = \frac{\sum_{i = 1}^{15} TP_i}{N_{segment}}\\
 \pi_e = \frac{\sum_{i = 1}^{15} \left\{(TP_i + FP_i) (TP_i + FN_i)\right\}}{{N_{segment}}^2}, \,\, \kappa = \frac{IR - \pi_e}{1 - \pi_e},
\end{eqnarray*}
where $N_{segment}$ is the total number of segments.}

\section{Results}
The biometric verification results within a one-to-one client-imposter classification problem are next summarised. In terms of the verification, we considered the following scenarios: 
\begin{itemize}
	\item Client-imposter verification based on varying segment lengths $L_{seg}$ (Section \ref{sec:results1}),
	\item Verification with various combinations of features (Section \ref{sec:results2}),
	\item Verification across different classifiers, both non-parametric and parametric ones (Section \ref{sec:results3}),
	\item Verification of registered clients and imposters ($S_R$), and of non-registered-imposters ($S_N$) (Section \ref{sec:results4}),
	\item Subject-wise verification (Section \ref{sec:results5}). 
\end{itemize}
We also considered biometric identification, that is, a one-to-many subject-to-subject classification problem (Section \ref{sec:results6}). {\color{black}Table \ref{table:summary} summarises the details of the considered scenarios. }

\begin{table*}[htbp]
\centering
\caption{{\color{black}Summary of parameter choice in the proposed biometric application}}
\label{table:summary}
\begin{tabular}{ccccccc}
 \hline
Section  & No. Subject& $L_{seg}$&Features&Classifier&Setup& System\\
\hline
\ref{sec:results1}  &15& $10,20,30,60,90$&PSD+AR&Cos. Dist.&R\&B & \multirow{5}{*}{Verification}\\
\ref{sec:results2}  &15& $60$&PSD, AR, PSD+AR&Cos. Dist.&R &\\
\ref{sec:results3}  & 15& $60$&PSD+AR&Cos. Dist., LDA, SVM&R\&B & \\
\ref{sec:results4}  & 15+5& $60$&PSD+AR&Cos. Dist., LDA, SVM&R\&B & \\
\ref{sec:results5}  & 15& $60$&PSD+AR&Cos. Dist.&R& \\
\cdashline{1-7}
\ref{sec:results6}  & 15& $10,20,30,60,90$ &PSD+AR&Cos. Dist.&R\&B& Identification\\
\hline
\end{tabular}
\end{table*}

\subsection{Client-Imposter verification with different segment sizes} \label{sec:results1}
Table \ref{table:Result_segment_size} summarises validation results for both Setup-R and Setup-B, over different segment sizes $L_{seg} = 10, 20, 30, 60, 90$ s. Both the PSD and AR features were used. The elements in TP, FN, FP and TN columns denote the number of segments classified by a validation stage. The chance level in this scenario is 14/15 = \unit[93.3]{\%}; this is because every subject is once used for client data, $V_{VC} \in \mathbb{R}^{N \times D}$ in Table \ref{table:data_structure}, and imposter data are selected from the other `non-client' subjects, $V_{VI} \in \mathbb{R}^{14N \times D}$ in Table \ref{table:data_structure}. The ratio between $V_{VC}$ and $V_{VI}$ is therefore $1:14$, and thus the chance level is 14/15.
In Setup-R, the results with \unit[$L_{seg} = 60$]{s} achieved both the best HTER score, \unit[17.2]{\%}, and the best accuracy (AC), \unit[95.7]{\%}. {\color{black} Notice that, the number of TP (=183) is larger than FN + FP (=174) with \unit[$L_{seg} = 60$]{s}, therefore, the likelihood of making a true positive verification is higher than making a false verification.}
In Setup-B, the results with \unit[$L_{seg} = 90$]{s} obtained both the best HTER score, \unit[6.9]{\%}, and the best accuracy (AC), \unit[98.3]{\%}.

\begin{table}[htbp]
\centering
\caption{Client-Impostor verification over different segment sizes}
\label{table:Result_segment_size}
\begin{tabular}{P{0.2cm}P{0.5cm}P{0.35cm}P{0.35cm}P{0.35cm} P{0.5cm}P{0.4cm}P{0.4cm}P{0.5cm}P{0.5cm}}
\hline
            &  $L_{seg}$&      TP& FN & FP& TN &  FAR & FRR &HTER& AC\\
 \hline
\multirow{5}{*}{\rot{\emph{\textbf{Setup-R}}}}
&10 s& 622  &       996   &      996  &     21656& 4.4 &61.6&33.0&91.8\\
&20 s&   371    &     439   &      439  &     10901 & 3.9 &54.2& 29.1&92.8\\
&30 s&   271   &      269  &       269  &      7291& 3.6 &49.8&26.7&93.4\\
&60 s&   183   &       87   &       87    &    3693&  {\bf 2.3} &{\bf 32.2}&{\bf 17.2 }& {\bf 95.7}\\
&90 s&   120   &       60   &       60   &     2460&  2.4 & 33.3& 17.8& 95.6\\
              \hline \hline
\multirow{5}{*}{\rot{\emph{\textbf{Setup-B}}}}
&10 s&1097  &       523     &    523  &     22157& 2.3&32.3&17.3&95.7\\
&20 s&   611   &      199    &     199  &     11141& 1.8&24.6& 13.2&96.7\\
&30 s&  425   &      115     &    115   &     7445   & 1.5 & 21.3&11.4&97.2\\
&60 s&   233   &       37    &      37   &     3743&  1.0&13.7& 7.3& 98.2\\
&90 s&  157    &       23   &       23   &     2497&  {\bf 0.9} & {\bf 12.8} &  {\bf 6.9}& {\bf 98.3}\\
\hline      
\multicolumn{10}{c}{TP: True Positive, FN: False Negative, FP: False Positive, TN: True} \\
\multicolumn{10}{c}{Negative, FAR: False Accept Rate, FRR: False Reject Rate,} \\
\multicolumn{10}{c}{HTER: Half Total Error Rate, AC: Accuracy} \\
              \hline
              \end{tabular}
\end{table}


\subsection{Client-Imposter verification with different features} \label{sec:results2}

Table \ref{table:Result_features} shows the validation results in Setup-R, and over a range of different selections of features, such as AR coefficients, frequency band power, and the combination of AR and band power features for the segment length of \unit[$L_{seg} = 60$]{s}. The classification results using both AR features and PSD features were the highest in terms of both HTER and AC, which corresponds to Table \ref{table:Result_segment_size} (upper-panel), for \unit[$L_{seg} = 60$]{s}.
\begin{table}[htbp]
\centering
\caption{Rigorous setup: Client-Imposter verification over different features in Setup-R}
\label{table:Result_features}
\begin{tabular}{cccccc}
\hline
  $L_{seg} = 60$            &No. feature $D$&        FAR & FRR  &HTER & AC \\
 \hline
AR &      20 & 4.8 &66.7&35.8&91.1\\
PSD &{\color{black}6}& 4.4&61.1& 32.8&91.9\\
AR + PSD&{\color{black}26}& {\bf 2.3} &{\bf 32.2}&{\bf 17.2}&{\bf 95.7}\\
              \hline
              \end{tabular}
\end{table}


\subsection{Client-Imposter verification with different classifiers} \label{sec:results3}
Table \ref{table:Result_segment_size_SVM} shows the imposter-client verification accuracy based on the minimum cosine distance, LDA, and SVM, for both Setup-R and Setup-B, with a segment size of \unit[$L_{seg} = 60$]{s}. Both the PSD and AR features were used. In Setup-R, the results with cosine distance were the best in terms of both HTER score, \unit[17.2]{\%} and AC, \unit[95.7]{\%}. 
In Setup-B, the results of both HTER and AC were the best based on the SVM classifier, \unit[5.5]{\%} and \unit[99.0]{\%}, respectively.

\begin{table}[htbp]
\centering
\caption{Client-Impostor verification with different classifiers}
\label{table:Result_segment_size_SVM}
\begin{tabular}{P{1.5cm}P{0.35cm}P{0.35cm}P{0.35cm} P{0.5cm}P{0.4cm}P{0.4cm}P{0.5cm}P{0.5cm}}
\hline
\unit[$L_{seg} = 60$]{s}&      TP& FN & FP& TN &  FAR & FRR &HTER& AC\\
 \hline
               \multicolumn{9}{c}{\emph{\textbf{Setup-R: rigorous}}} \\
              \hdashline
Cos. Dist &   183   &       87   &       87    &    3693&   2.3 &{\bf 32.2}&{\bf 17.2 }& {\bf 95.7}\\
LDA & 149    &     121      &   139  &      3641& 3.7& 44.8 & 24.2 & 93.6\\
SVM&  135  &       135          &54   &     3726& {\bf 1.4}  & 50.0 & 25.7& 95.3\\
              \hline \hline
              \multicolumn{9}{c}{\emph{\textbf{Setup-B: biased}}} \\
              \hdashline
Cos. Dist &  233   &       37    &      37   &     3743&  1.0& 13.7&7.3& 98.2\\ 
LDA&   200      &    70    &      87  &      3693&  2.3&25.9&14.1&96.1\\ 
SVM&   241    &      29  & 11   &     3769  & {\bf 0.3}&{\bf 10.7}& {\bf 5.5}& {\bf 99.0}\\
\hline      
              \end{tabular}
\end{table}


\subsection{Validation including non-registered imposters} \label{sec:results4}
Table \ref{table:Result_which_dataset} summarises the confusion matrices of both Setup-R and Setup-B with segment sizes \unit[$L_{seg} = 60$]{s}, classified by the minimum cosine distance, LDA, and SVM; these correspond to Table \ref{table:Result_segment_size_SVM}, panels Setup-R and Setup-B. The confusion matrices were categorised into:
\begin{itemize}
	\item Client matrix $Y_{VC}$ from dataset $S_R$,
	\item Imposter matrix $Y_{VI}$ from dataset $S_R$
	\item Imposter matrix $Y_{VI_{N}}$ from dataset $S_N$. 
\end{itemize}
{\color{black} Notice that the minimum cosine distance approach assigns the class (client or imposter) of the nearest data from the training matrix. In this study, we selected every trial from every subject for the validation of client data, so as to have validated ninety times; therefore, \emph{the nearest data (from dataset $S_R$) for each imposter data from dataset $S_N$, also always become the `client' once.} Hence, when the nearest data for an $S_N$ data become the `client' data in the training matrix, the imposter data from dataset $S_N$ are straightforwardly classified as `client'. Therefore, regardless of data, the TPR for Imposter matrix $Y_{VI_{N}}$ is \unit[93.3]{\%} for the minimum cosine distance approach; however for comparison among different classifiers, these results are also included.}

In Setup-R, the TPR of client $Y_{VC}$, achieved by the minimum cosine distance was the highest, with respective value of \unit[67.8]{\%}. However, the TPRs obtained by SVM for imposters $Y_{VI}$ and $Y_{VI_{N}}$ were \unit[98.6]{\%} and \unit[96.2]{\%}, respectively, which was higher than those achieved by LDA. 
In Setup-B, both the TPR of client $Y_{VC}$ and that of imposters $Y_{VI}$ and $Y_{VI_{N}}$ by SVM were the highest, with respective values of \unit[89.3]{\%}, \unit[99.7]{\%} and \unit[96.3]{\%}.

\begin{table}[htbp]
\centering
\caption{Confusion matrix of the Client-Imposter verification scenario from different datasets in both Setup-R and Setup-B}
\label{table:Result_which_dataset}
\begin{tabular}{P{0.1cm} P{0.1cm} P{3cm} P{0.6cm}  P{0.7cm}  P{0.7cm} | P{0.7cm}}
\hline
 \multicolumn{3}{c}{\multirow{2}{*}{\unit[$L_{seg} = 60$]{s}}}& \multicolumn{2}{c}{Prediction} & & TPR\\
   										&&& Client  & Imposter & Total  & (\%)\\
 \hline
       \multirow{9}{*}{\rot{\emph{\textbf{Setup-R}} Label}}  &\multirow{3}{*}{\rot{Cos.}} &
       Client $Y_{VC}$&      183 & 87 & 270 &{\bf 67.8}\\
     & &Imposter $Y_{VI}$&    87 &  3693 &3780&97.7\\
      & &Imposter $Y_{VI_{N}}$$^\ast$  &90$^\ast$& 1260$^\ast$&1350$^\ast$& 93.3$^\ast$\\
              \cdashline{2-7}
          &  \multirow{3}{*}{\rot{LDA}}  &
          Client $Y_{VC}$&149    &     121&270&55.2\\
          &&Imposter $Y_{VI}$           &139   &     3641 & 3780 & 96.3\\
         &  &Imposter $Y_{VI_{N}}$           &136    &    1214 & 1350 & 89.9\\
              \cdashline{2-7}
          &  \multirow{3}{*}{\rot{SVM}}  &
          Client $Y_{VC}$&135 &  135&270&50.0\\
          &&Imposter $Y_{VI}$           &54 &  3726& 3780 & {\bf 98.6}\\
         &  &Imposter $Y_{VI_{N}}$           & 52 &  1298 & 1350 &{\bf 96.2}\\
\hline \hline
       \multirow{9}{*}{\rot{\emph{\textbf{Setup-B}} Label}}& \multirow{3}{*}{\rot{Cos.}} &
       Client $Y_{VC}$&      233    &     37 & 270 &86.3\\
      & &Imposter $Y_{VI}$&    37   &     3743 &3780&99.0\\
      & &Imposter $Y_{VI_{N}}$$^\ast$ &90$^\ast$& 1260$^\ast$& 1350$^\ast$& 93.3$^\ast$\\
               \cdashline{2-7}
          &   \multirow{3}{*}{\rot{LDA}} &
       Client $Y_{VC}$&      200    &      70 & 270 &74.1\\
      & &Imposter $Y_{VI}$&     87    &    3693 &3780&97.7\\
      & &Imposter $Y_{VI_{N}}$  &152   &     1198 & 1350& 88.7\\
          \cdashline{2-7}
          &  \multirow{3}{*}{\rot{SVM}}  &
          Client $Y_{VC}$&241&   29&270&{\bf 89.3}\\
          &&Imposter $Y_{VI}$           &11 &  3769& 3780 & {\bf 99.7}\\
         &  &Imposter $Y_{VI_{N}}$           & 50 &  1300 & 1350 &{\bf 96.3}\\
              \hline
              \multicolumn{7}{c}{$^\ast$ Always the same result regardless of the data} \\
              \hline
              \end{tabular}
\end{table}

\subsection{Client-Imposter verification results per subject} \label{sec:results5}
Table \ref{table:Result_subjects} (middle columns) summarises the subject- and day-wise validation results with PSD and AR features from $L_{seg} = $ \unit[60]{s} segments in Setup-R, which corresponds to Table \ref{table:Result_segment_size} (upper-panel) for $L_{seg} = $ \unit[60]{s}. The first and second columns in the Verification part show respectively subject-wise HTER and AC, with the training matrix $Y_T$ selected from all the first day recordings and classified based on the second day recordings. The third and fourth columns in the Verification part show classification results obtained based on the training matrix $Y_T$, which was selected from all the second day recordings in order to classify the first day recordings.


\subsection{Biometrics identification scenarios} \label{sec:results6}
\begin{figure}[tb]
\begin{center}
    		\includegraphics[clip, width=\linewidth]{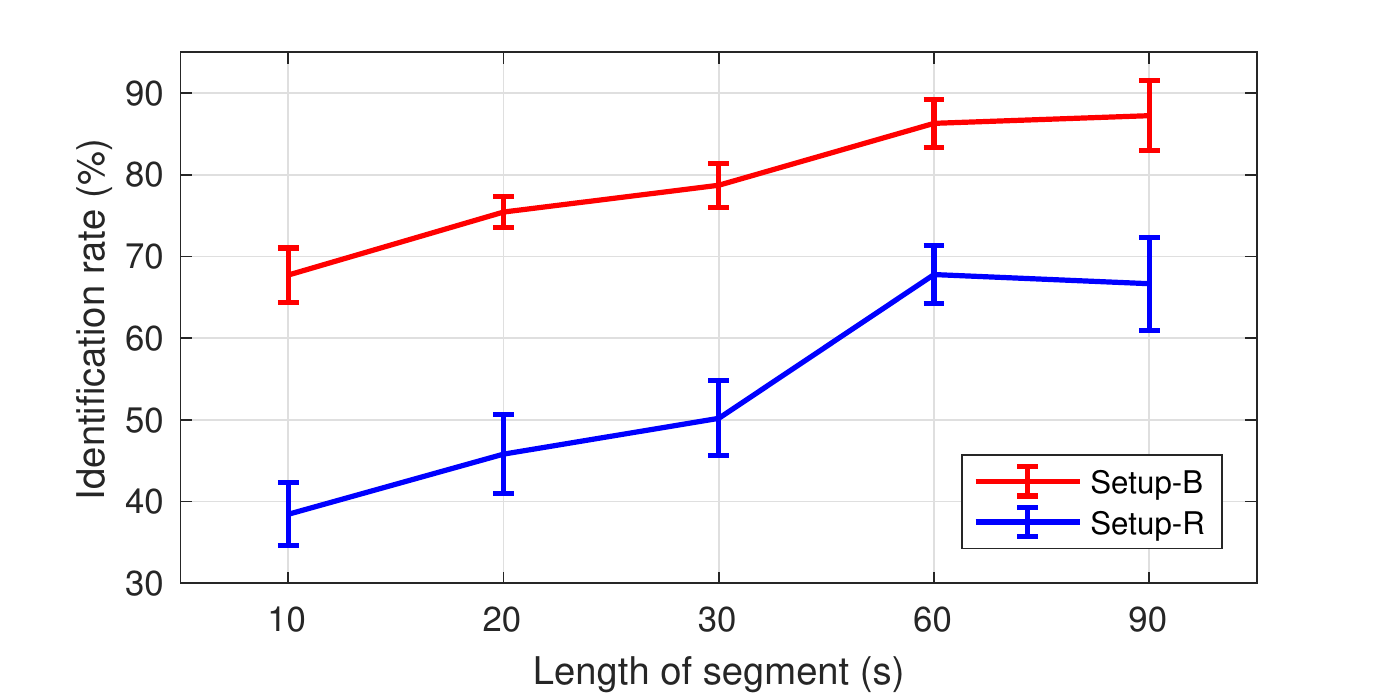}
    	\caption{{\color{black}The identification rate with different segment size in Setup-R and Setup-B. The error bars indicate the standard error.}}
	\label{fig:identification_rate}
\end{center}
	\end{figure}

Table \ref{table:Result_subjects} (right column) summarises the subject-wise identification rate obtained by the minimum cosine distance classifier with the PSD and AR features from \unit[$L_{seg} =  60$]{s} segments in Setup-R. Previously, we considered a binary client-imposter classification problem (e.g. $M=2$) for each subject, each day, and each trial, however, the classification algorithm used in this study was the simple minimum cosine distance between the validation matrix and training matrix. 
For the prediction of $l$th row of the validation matrix, $Y_{V_{(l,:)}}$, the minimum distance between the training matrix, $Y_{T}$, was found, e.g. the $n$th row of the training matrix, and the same label was assigned to the $n$th row of the training matrix as the prediction label for $l$th row of the validation matrix, $Y_{V_{(l,:)}}$. 
Notice that \emph{the minimum distance approach is applicable for biometrics identification problems, which is a one-to-many classification. } {\color{black} The number of classes was $M = 15$, which corresponds to the the number of subjects in the $S_R$ recordings, therefore the chance level was 1/15 = \unit[6.7]{\%}. The achieved identification rate was \unit[67.8]{\%} with \unit[$L_{seg} =  60$]{s} segments in Setup-R, while the achieved Cohen's kappa coefficient was $\kappa = 0.65$ (Substantial agreement) \cite{Landis2008}.

Figure \ref{fig:identification_rate} shows identification rate of both Setup-R and Setup-B, with different segment sizes $L_{seg} = 10, 20, 30, 60, 90$ s. In Setup-B, the identification rate with $L_{seg} = 10$ s was \unit[67.7]{\%}, which was almost the same as the result with $L_{seg} = 60$ s in Setup-R. The highest identification rate, \unit[87.2]{\%}, was achieved with $L_{seg} = 90$ s, where corresponding Kappa was $\kappa = 0.86$ (Almost Perfect agreement) in Setup-B.}

\begin{table*}[htbp]
\centering
\caption{\color{black} Recording details, accuracy and HTER in the verification problem, and sensitivity in the identification problem for each subject with different training and validation data in Setup-R}
\label{table:Result_subjects}
\begin{tabular}{cccccc|cccc|c}
\hline
        & &    \multicolumn{2}{c}{Impedance}     & &  & \multicolumn{4}{c|}{{\bf Verification} (Client-Imposter, $M=2$)} & {\bf Identification} \\
    $S_R$    & Interval &  \multicolumn{2}{c}{(k$\Omega$)}  &   \multicolumn{2}{c|}{Time$^\ast$}  & \multicolumn{2}{c}{$Y_T$:Day1 $\rightarrow$ $Y_V$:Day2} & \multicolumn{2}{c|}{$Y_T$:Day2 $\rightarrow$ $Y_V$:Day1} & ($M=15$) \\
      Sub&day&Day1&Day2&Day1&Day2&AC(\%) & HTER(\%)  &AC(\%) &HTER(\%) & SE(\%)\\
 \hline
1  & 5&    $< 9$&$< 10$&M&A& {\bf 98.5}& 6.0 & {\bf 100} & {\bf 0.0} & {\bf 94.4}\\
2  &  6&   $< 6$&$< 6$&A&A&  94.8 & 28.6 & 97.0 & 11.9 & 61.1\\
3 & 8&   $< 5$&$< 8$&A&A& {\bf 98.5} & {\bf 0.8} & 94.8 & 33.8 & 66.7\\
4 &  7 &  $< 9$&$< 10$&A&M&  95.6 & 28.2 & 96.3 & 7.2 & 66.7\\
5 &   7&  $< 8$&$< 4$&A&A&  96.3 & 17.5 & 97.0 & 22.2 & 61.1\\
6 & 7& $< 12$&$< 14$&A&A&{\it 91.1} & 20.2 & 97.8 & 6.4 & 77.8\\
7 & 15 & $< 12$&$< 13 $&M&M&{\bf 98.5} & 11.1 & 96.3 & 22.6 & 66.7\\
8 & 6 & $< 11$&$< 11 $&M&M&{\it 91.1}& {\it 35.8} &{\it 91.9}&{\it 35.4}& {\it 33.3}\\
9 & 7 & $< 10$&$<13 $&A&A&96.3 & 12.3 & 95.6 & 12.7 & 77.8  \\
10  & 8 & $< 9$&$< 9 $&M&M& 91.9 & 35.4 & 93.3 & 8.7 & 61.1 \\
11 & 6 & $< 11$&$< 13$&A&A& 95.6 & 17.8 & 94.8 & 13.1 & 72.2\\
12  & 5 & $< 13$&$< 11$&M&M& 96.3 & 17.5 & 99.3 & 0.4 & 83.3\\
13  & 7 & $< 13$&$<9 $&M&M& 95.6 & 17.8 & 97.8 & 11.5 & 72.2 \\
14  & 6 & $< 8$&$< 9$&M&A& 97.8 & 11.5 & 95.6 & 33.4 & 55.6\\
15  & 5 & $< 9$&$< 13$&A&A&94.1 & 13.5 &{\it 91.9}& 25.0 & 66.7 \\
\hline
Ave.   &7 &  - & - &-&-&95.5&18.2 &96.0 & 16.3 & IR = \unit[{\bf 67.8}]{\%}\\
\hline
\multicolumn{6}{c}{\emph{\textbf{Setup-R: rigorous}}, $L_{seg} = 60$}& \multicolumn{4}{c}{Overall AC=\unit[95.7]{\%}, HTER=\unit[17.2]{\%}} &$\kappa$ = 0.65\\
              \multicolumn{6}{c}{$^\ast$M: Morning (9-12), A: Afternoon (12-18)}& \multicolumn{5}{c}{{\bf bold}: best result, {\it italic}: worst result, out of 15 subjects.}\\
              \hline
              \end{tabular}
\end{table*}

\section{Discussion}

This study aims to establish a repeatable and highly collectable EEG biometrics using a wearable in-ear sensor. We considered a biometric verification problem, which was cast into a  one-to-one client-imposter classification setting. Notice that, as described in Section \ref{sec:classification}, before classification, the validation matrix was normalised column-wise to the range $[0, 1]$ using the corresponding maximum/minimum values of the training matrix. 

\subsection{Verification with different segment sizes}
{\color{black} Firstly, the classification results were compared for different segment lengths $L_{seg}$, shown in Table \ref{table:Result_segment_size}. Within the same setup, i.e. Setup-R or Setup-B, the performance of HTER and AC increased with the segment length, although the results with $L_{seg} = 60$ s and  $L_{seg} = 90$ s are almost the same. 
Longer segments allowed for more data epochs to be averaged over, hence the EEG noise inference for classification diminished and the inherent EEG characteristic were able to be captured by averaging. However, a longer segment length also implies a longer recording time, which is not ideal for feasible EEG biometrics. 
Compared to the results in Setup-R and Setup-B for the same segment length, both the HTERs and accuracy (ACs) of Setup-B were clearly better than those of Setup-R.
In terms of client discrimination, the decrease in FRR was significant from Setup-R to Setup-B. In Setup-B, a larger number of client segments was correctly classified (see TP) than in Setup-R. 
In setup-R, only the result with \unit[$L_{seg} = 60$]{s} achieved TP > (FN + FP), which indicates that the likelihood of making a true positive verification is higher than making a false verification; in Setup-B, the shortest segment size, \unit[$L_{seg} = 10$]{s}, achieved TP > (FN + FP).

The difference between Setup-R and Setup-B was that the training matrices $Y_{T}$ in Setup-B included the trials which were recorded on the same day as the validation trial. In other words, the assigned validation data (trial) and the part of assigned training data (trials) were recorded within 5 - 10 minutes in the same environment. Therefore, the training matrix contains significantly similar EEG recordings to the validation matrix in Setup-B, and this leads to a higher classification performance than the classification in Setup-B. 

With an increase in the segment size $L_{seg}$, the number of segments per recording trial, $N$, became smaller, especially for $N = 2$ with the segment size $L_{seg} = 90$ s; therefore, the training matrix only contains six (2 $\times$ 3 trials) examples of client data in Setup-R ({\it c.f.} ten examples in Setup-B), which might not be enough for training client data. Hence, the performance with $L_{seg} = 90$ s was slightly lower than that with $L_{seg} = 60$ s in Setup-R.}
%


\subsection{Verification with different classifiers}
{\color{black} Table \ref{table:Result_segment_size_SVM} shows the classification comparison among the minimum cosine distance methods, LDA and SVM. The SVM was used as a parametric classifier; 
firstly, the optimal hyper-parameters (see details in Table \ref{table:kernels}) were selected from 5-fold cross-validation within the training matrix, and then weight parameters based on these chosen hyper-parameters were obtained. 
Notice that we could tune the classifier in different ways, e.g. in order to minimise false acceptance or minimise false rejection. {\it The optimal tuning in this study was performed so as to maximise class sensitivities, i.e. maximise the number of TP and TN elements, which resulted in minimum HTERs.} In both Setup-R and Setup-B, the FARs by SVM were smaller than those achieved by both the minimum cosine distance and the LDA, because the tuning was performed for maximising TN elements. Since the number of imposter elements was fourteen times bigger than the number of clients in both Setup-R and Setup-B (i.e. chance level was 14/15), the SVM parameters were tuned for higher sensitivity to imposters. As a result, the FRR by SVM, which were related to client sensitivity given in Table \ref{table:Result_which_dataset}, were higher than those achieved by both minimum cosine distance and LDA in Setup-R. 

In Setup-B, as mentioned above, the training matrix contains the data from the same recording day, which are more similar EEG patterns than the data obtained from a different recording day. Therefore, the SVM model chose hyper-parameters and weight parameters from the training matrix, so as to better the validation data in Setup-B, which led to higher performance than by both the minimum cosine distance and LDA.}

Notice that, as described before, tuning of the hyper-parameters was performed within the training matrix, then the so-obtained hyper-parameters were used for finding the optimal weight parameters within the training matrix. The same hyper-parameters and weight parameters were used for classifying the validation matrix. This setup is applicable for feasible EEG biometrics scenarios in the real-world. 

\subsection{Validation including non-registered imposters}
{\color{black} In Table \ref{table:Result_which_dataset}, the confusion matrices for the client matrix $Y_{VC}$ and imposter matrix $Y_{VI}$ from dataset $S_R$ and imposter matrix $Y_{VI_{N}}$, are given, which were then used for a comparison between Setup-R and Setup-B. 
Compared to the results obtained within the same classifier (minimum cosine distance, LDA, SVM) in Setup-R and Setup-B, the true positive rate (TPR) of clients $Y_{VC}$ and the sensitivity of imposter  $Y_{VI}$ from dataset $S_R$ in Setup-B were higher than those in Setup-R. As described before, in Setup-B, the two client trials from the same recording day as the validation trial were included into the training matrix, and therefore more segments were correctly classified. 
In contrast, the TPRs of imposter $Y_{VI_{N}}$ from dataset $S_N$ by both the LDA and SVM in Setup-B were almost the same to those in Setup-R; \unit[89.9]{\%} and \unit[88.7]{\%} for LDA, \unit[96.2]{\%} and \unit[96.3]{\%} for SVM. Compared to the TPR between two imposter data ($Y_{VI}$ and $Y_{VI_{N}}$), regardless of the classifiers, the TPR of $Y_{VI}$ were higher than those of $Y_{VI_{N}}$. Since the imposter data from $S_N$ were not included in the training matrix, the more $S_N$ data were misclassified as `client' than $S_R$ data misclassified as `client'. 

However, in the real-world scenarios for biometrics, imposters are not always `registered'. The lower TPR for $Y_{VI_{N}}$ means that the application is inadequate for attack from non-registered subjects. One potential way to overcome the vulnerability of the minimum cosine distance classifier is by introducing threshold for classification. If the nearest distance is larger than the given distance parameter, the segment is excluded from the classification or is classified as imposter.}


\subsection{Client-Imposter verification results per subject}
\begin{figure}[tb]
\begin{center}
    		\includegraphics[clip, width=\linewidth]{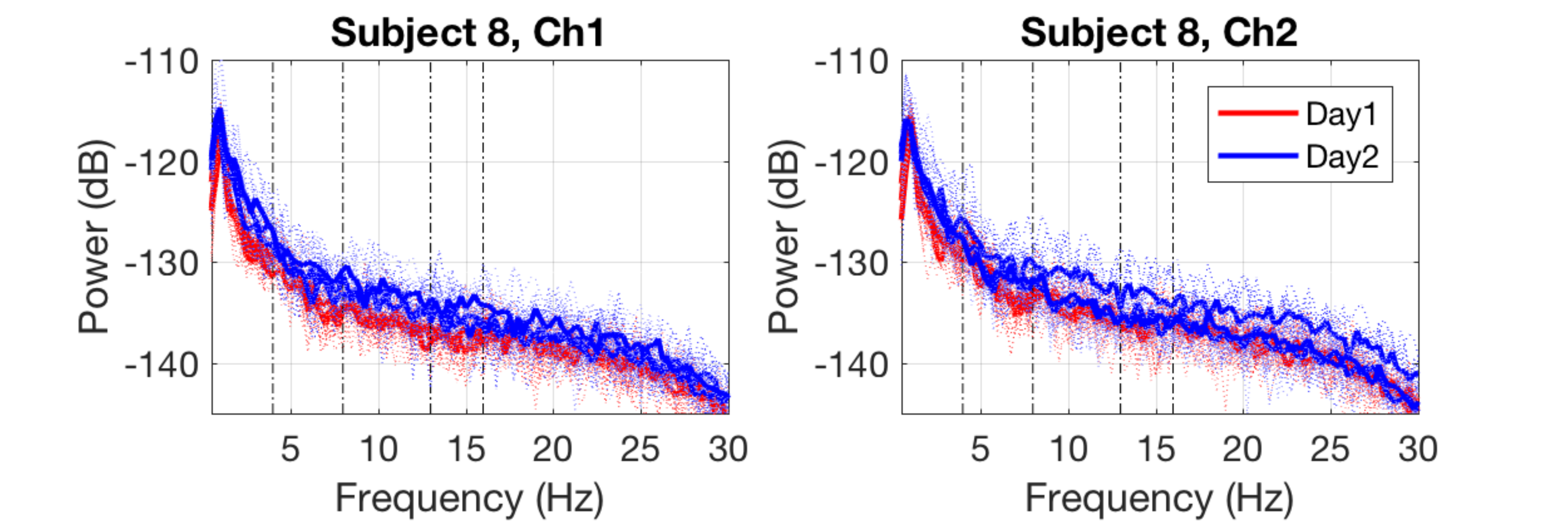}
    	\caption{{\color{black} Power spectral density for the in-ear EEG Ch1 (left) and the in-ear EEG Ch2 (right) of Subject 8. The thick lines correspond to the averaged periodogram obtained by the all recordings from the 1st day (red) and the 2nd day (blue), whereas the thin lines are the averaged periodogram obtained by a single trial.}}
	\label{fig:psd_sample2}
\end{center}
	\end{figure}

{\color{black} For subject-wise classification, Table \ref{table:Result_subjects} summarises classification results obtained by the minimum cosine distance in Setup-R, for different training-validation scenarios. The results varied across subjects and for training-validation configurations between \unit[91.1]{\%} to \unit[100]{\%} of AC and between \unit[0.0]{\%} to \unit[35.8]{\%} of HTER. 

The size of viscoelastic earpiece was the same for twenty subjects (both $S_R$ and $S_N$ subjects), therefore all the subjects were able to wear it comfortably. The upper bounds of the electrode impedance over three recordings per day of each participant are given in Table \ref{table:Result_subjects} (Impedance part). The highest performance was achieved by Subject 1, with maximum impedances of 9 k$\Omega$ and 10 k$\Omega$ for the 1st and 2nd recording, respectively. Even though the impedances for Subject 2 and Subject 5 were smaller than those for Subject 1, the corresponding performance was below average over fifteen subjects. Besides, the lowest performance was exhibited by Subject 8, for whom the impedances were smaller than 11 k$\Omega$ for Day1 and 11 k$\Omega$ for Day2. Figure \ref{fig:psd_sample2} shows average PSDs for Subject 8 -- observe that the PSDs for the EEG recorded on Day2 (blue) is slightly larger than those on Day1 (red).}

\subsection{Biometrics identification}
{\color{black} In terms of biometrics identification results, a one-to-many subject-to-subject classification problem, the average sensitivity over fifteen subjects, i.e. the identification rate, was \unit[67.8]{\%} in Setup-R with $L_{seg} = 60$ s, as shown in Table \ref{table:Result_subjects} (right column). Figure \ref{fig:identification_rate} illustrates the identification rates of both Setup-R and Setup-B, with different segment sizes $L_{seg} = 10, 20, 30, 60, 90$ s. The identification rate increased with segment length, although the results with $L_{seg} = 60$ s and  $L_{seg} = 90$ s are almost the same.

Notice that the performances with $L_{seg} = 10$ s in Setup-B and $L_{seg} = 60$ s in Setup-R were almost the same, \unit[67.7]{\%} and \unit[67.8]{\%}, respectively. The highest identification rate in Setup-B was \unit[87.2]{\%} with $L_{seg} = 90$ s. Indeed, the training matrix for Setup-B included the trials which were recorded at the same day as the validation trial; therefore the performance was better than in Setup-R.

In a previous biometrics identification study, Maiorana {\it et al.} \cite{Maiorana2016a} analysed 19 channels of EEG during EC tasks in three different recording days, and achieved the rank-1 identification rate (R1IR) of \unit[90.8]{\%} for a segment length \unit[45]{s}. Notice that it is hard to compare the performance with our approach, because the number of channels was very different, as 19 scalp EEG channels covered the entire head vs our 2 in-ear EEG channels embedded on an earplug. Therefore, although our results were lower, the proof-of-concept in-ear biometrics emphasised the \emph{collectability} aspect in fully wearable scenarios. }

\subsection{Alpha attenuation in the real-world scenarios}
\begin{figure}[tb]
\begin{center}
    		\includegraphics[clip, width=\linewidth]{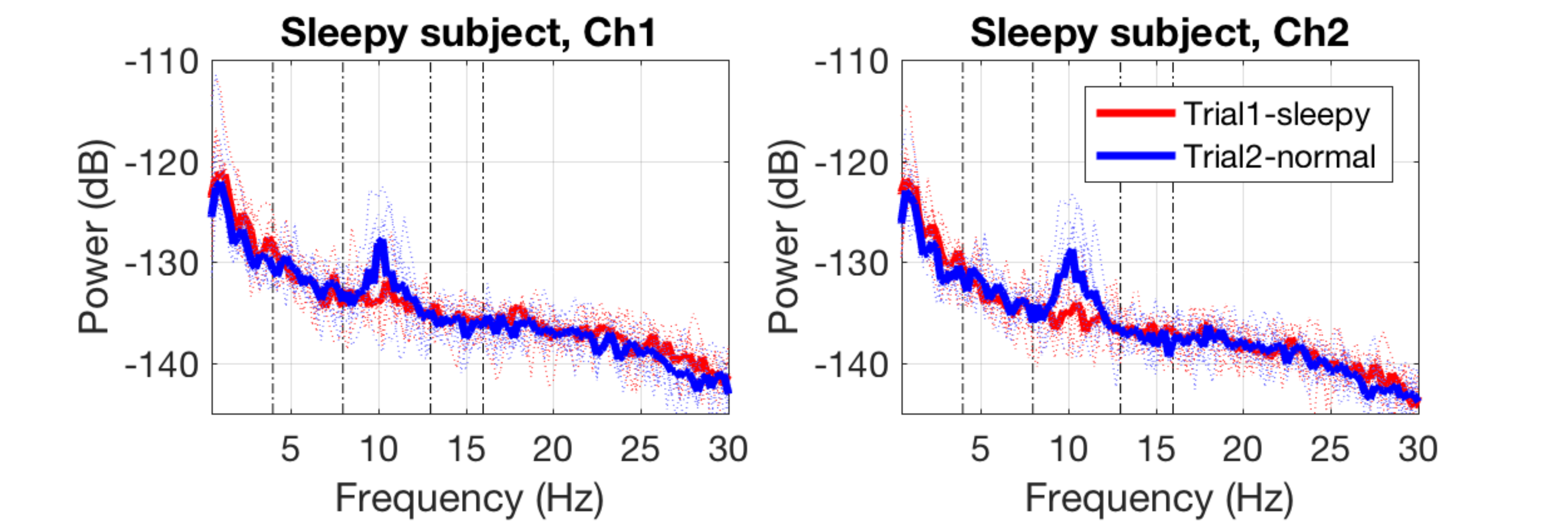}
    	\caption{{\color{black} Power spectral density for the in-ear EEG Ch1 (left), and the in-ear EEG Ch2 (right) of one subject. The thick lines correspond to the averaged periodogram obtained by the recordings from `sleepy' trial (red) and `normal' trial (blue). Observe that the alpha power attenuated during the `sleepy' trial.}}
	\label{fig:psd_sleepy}
\end{center}
	\end{figure}

{\color{black} One limitation of using the alpha band, is the sensitivity to drowsiness, a state where the alpha band power is naturally elevated. For illustration, Figure \ref{fig:psd_sleepy} shows the PSD obtained from a subject, calculated by Welch's averaging periodogram method. The subject slept during one recording, then the subject was woken up and another recording started less than 10 minutes after the first recording. The PSD graphs in Figure \ref{fig:psd_sleepy} are overlapped except for the alpha band; the alpha power observed during the `sleepy' recording trial was smaller than that at the `normal' recording, thus demonstrating the alpha attenuation due to fatigue, sleepiness, and drowsiness. The alpha attenuation is well known in the research in sleep medicine \cite{Silber2007, Nakamura2017a}, where it is particularly used to monitor sleep onset.}

\section{Conclusion}
We have introduced a proof-of-concept for a feasible, collectable and reproducible EEG biometrics in the community by virtue of an unobtrusive, discreet, and convenient to use in-ear EEG device. We have employed robust PSD and AR features to identify an individual, and unlike most of the existing studies, we have performed classification rigorously, without mixing the training and validation data from the same recording days. We have achieved HTER of \unit[17.2]{\%} with AC of \unit[95.7]{\%} with segment sizes of \unit[60]{s}, over the dataset from fifteen subjects. 

{\color{black} The aspects that need to be further addressed in order to fulfil the requirements for `truly wearable biometrics' in the `real-world' will focus on extensions and generalisations of this proof-of-concept to cater for:
\begin{itemize}
	\item Intra-subject variability with respect to the circadian cycle and the mental state, such as fatigue, sleepiness, and drowsiness;
	\item Additional feasible recording paradigms, for example, evoked response scenarios;
	\item Truly wearable scenarios with mobile and affordable amplifiers;
	\item Inter- and intra-subject variability over the period of months and years;
	\item Fine tuning of the variables involved in order to identify the optimal features and parameters (segment length, additional EEG bands). 
\end{itemize}
}

\section{Acknowledgement}
We wish to thank the anonymous reviewers for their insightful comments.

\bibliographystyle{ieeetr}
{\small
\bibliography{library}

\begin{thebibliography}{10}

\bibitem{Jain2000}
A.~Jain, L.~Hong, and S.~Pankanti, ``{Biometric identification},'' {\em
  Communications of the ACM}, vol.~43, no.~2, pp.~90--98, 2000.

\bibitem{Sato2013}
Y.~Sato, F.~Akazawa, D.~Muramatsu, T.~Matsumoto, A.~Nakamura, and T.~Sota,
  ``{An authentication method by high spectral resolution palm datacube},'' in
  {\em Proceedings of the International Conference on Biometrics and Kansei
  Engineering}, pp.~239--244, 2013.

\bibitem{Holland2013}
C.~D. Holland and O.~V. Komogortsev, ``{Complex eye movement pattern
  biometrics: The effects of environment and stimulus},'' {\em IEEE
  Transactions on Information Forensics and Security}, vol.~8, no.~12,
  pp.~2115--2126, 2013.

\bibitem{Odinaka2010}
I.~Odinaka, P.~H. Lai, A.~D. Kaplan, J.~A. O'Sullivan, E.~J. Sirevaag, S.~D.
  Kristjansson, A.~K. Sheffield, and J.~W. Rohrbaugh, ``{ECG biometrics: A
  robust short-time frequency analysis},'' in {\em Proceedings of IEEE
  International Workshop on Information Forensics and Security (WIFS)}, 2010.

\bibitem{Liu2014}
Y.~Liu and D.~Hatzinakos, ``{Earprint: Transient evoked otoacoustic emission
  for biometrics},'' {\em IEEE Transactions on Information Forensics and
  Security}, vol.~9, no.~12, pp.~2291--2301, 2014.

\bibitem{Prabhakar2003}
S.~Prabhakar, S.~Pankanti, and A.~Jain, ``{Biometric recognition: Security and
  privacy concerns},'' {\em IEEE Security {\&} Privacy Magazine}, vol.~1,
  no.~2, pp.~33--42, 2003.

\bibitem{Wolpaw2002}
J.~R. Wolpaw, N.~Birbaumer, D.~J. McFarland, G.~Pfurtscheller, and T.~M.
  Vaughan, ``{Brain computer interfaces for communication and control},'' {\em
  Frontiers in Neuroscience}, vol.~4, no.~113, pp.~767--791, 2002.

\bibitem{Johnson1959}
L.~C. Johnson and G.~A. Ulett, ``{Quantitative study of pattern and stability
  of resting electroencephalographic activity in a young adult group},'' {\em
  Electroencephalography and Clinical Neurophysiology}, vol.~11, no.~2,
  pp.~233--249, 1959.

\bibitem{Berkhout1968}
J.~Berkhout and D.~O. Walter, ``{Temporal stability and individual differences
  in the human EEG: An analysis of variance of spectral values},'' {\em IEEE
  Transactions on Biomedical Engineering}, no.~3, pp.~165--168, 1968.

\bibitem{Palaniappan2007a}
R.~Palaniappan and D.~P. Mandic, ``{EEG based biometric framework for automatic
  identity verification},'' {\em Journal of VLSI Signal Processing}, vol.~49,
  no.~2, pp.~243--250, 2007.

\bibitem{Palaniappan2007}
R.~Palaniappan and D.~P. Mandic, ``{Biometrics from brain electrical activity:
  A machine learning approach},'' {\em IEEE Transactions on Pattern Analysis
  and Machine Intelligence}, vol.~29, no.~4, pp.~738--742, 2007.

\bibitem{Looney2012}
D.~Looney, P.~Kidmose, C.~Park, M.~Ungstrup, M.~Rank, K.~Rosenkranz, and D.~P.
  Mandic, ``{The in-the-ear recording concept: User-centered and wearable brain
  monitoring},'' {\em IEEE Pulse}, vol.~3, no.~6, pp.~32--42, 2012.

\bibitem{Campisi2014}
P.~Campisi and D.~L. Rocca, ``{Brain waves for automatic biometric-based user
  recognition},'' {\em IEEE Transactions on Information Forensics and
  Security}, vol.~9, no.~5, pp.~782--800, 2014.

\bibitem{Rocca2014}
D.~L. Rocca, P.~Campisi, and G.~Scarano, ``{Stable EEG features for biometric
  recognition in resting state conditions},'' {\em Biomedical Engineering
  Systems and Technologies}, vol.~452, pp.~313--330, 2014.

\bibitem{Abdullah2010}
M.~K. Abdullah, K.~S. Subari, J.~L.~C. Loong, and N.~N. Ahmad, ``{Analysis of
  effective channel placement for an EEG-based biometric system},'' in {\em
  Proceedings of IEEE EMBS Conference on Biomedical Engineering and Sciences
  (IECBES)}, pp.~303--306, 2010.

\bibitem{Riera2008}
A.~Riera, A.~Soria-Frisch, M.~Caparrini, C.~Grau, and G.~Ruffini,
  ``{Unobtrusive biometric system based on electroencephalogram analysis},''
  {\em EURASIP Journal on Advances in Signal Processing}, vol.~2008, 2008.

\bibitem{Su2010}
F.~Su, L.~Xia, A.~Cai, Y.~Wu, and J.~Ma, ``{EEG-based personal identification:
  From proof-of-concept to a practical system},'' in {\em Proceedings of
  International Conference on Pattern Recognition}, pp.~3728--3731, 2010.

\bibitem{Marcel2007}
S.~Marcel and J.~d.~R. Millan, ``{Person authentication using brainwaves (EEG)
  and maximum a posteriori model adaptation},'' {\em IEEE Transactions on
  Pattern Analysis and Machine Intelligence}, vol.~29, no.~4, pp.~743--748,
  2007.

\bibitem{Lee2013}
H.~J. Lee, H.~S. Kim, and K.~S. Park, ``{A study on the reproducibility of
  biometric authentication based on electroencephalogram (EEG)},'' in {\em
  Proceedings of the IEEE International EMBS Conference on Neural Engineering
  (NER)}, pp.~13--16, 2013.

\bibitem{Rocca2013}
D.~L. Rocca, P.~Campisi, and G.~Scarano, ``{On the repeatability of EEG
  features in a biometric recognition framework using a resting state
  protocol},'' in {\em Proceedings of the International Conference on
  Bio-inspired Systems and Signal Processing (BIOSIGNALS)}, pp.~419--428, 2013.

\bibitem{Armstrong2015}
B.~C. Armstrong, M.~V. Ruiz-Blondet, N.~Khalifian, K.~J. Kurtz, Z.~Jin, and
  S.~Laszlo, ``{Brainprint: Assessing the uniqueness, collectability, and
  permanence of a novel method for ERP biometrics},'' {\em Neurocomputing},
  vol.~166, pp.~59--67, 2015.

\bibitem{Maiorana2016a}
E.~Maiorana, D.~{La Rocca}, and P.~Campisi, ``{On the permanence of EEG signals
  for biometric recognition},'' {\em IEEE Transactions on Information Forensics
  and Security}, vol.~11, no.~1, pp.~163--175, 2016.

\bibitem{Chen2016}
Y.~Chen, A.~D. Atnafu, I.~Schlattner, W.~T. Weldtsadik, M.~C. Roh, H.~J. Kim,
  S.~W. Lee, B.~Blankertz, and S.~Fazli, ``{A high-security EEG-based login
  system with RSVP stimuli and dry electrodes},'' {\em IEEE Transactions on
  Information Forensics and Security}, vol.~11, no.~12, pp.~2635--2647, 2016.

\bibitem{Kidmose2013a}
P.~Kidmose, D.~Looney, M.~Ungstrup, M.~L. Rank, and D.~P. Mandic, ``{A study of
  evoked potentials from ear-EEG},'' {\em IEEE Transactions on Biomedical
  Engineering}, vol.~60, no.~10, pp.~2824--2830, 2013.

\bibitem{Looney2016a}
D.~Looney, V.~Goverdovsky, I.~Rosenzweig, M.~J. Morrell, and D.~P. Mandic, ``{A
  wearable in-ear encephalography sensor for monitoring sleep: Preliminary
  observations from nap studies},'' {\em Annals of the American Thoracic
  Society}, vol.~13, no.~12, pp.~2229--2233, 2016.

\bibitem{Nakamura2017a}
T.~Nakamura, V.~Goverdovsky, M.~J. Morrell, and D.~P. Mandic, ``{Automatic
  sleep monitoring using ear-EEG},'' {\em IEEE Journal of Translational
  Engineering in Health and Medicine}, vol.~5, no.~1, p.~2800108, 2017.

\bibitem{Goverdovsky2017}
V.~Goverdovsky, W.~von Rosenberg, T.~Nakamura, D.~Looney, D.~J. Sharp,
  C.~Papavassiliou, M.~J. Morrell, and D.~P. Mandic, ``{Hearables: Multimodal
  physiological in-ear sensing},'' {\em Scientific Reports}, vol.~7, no.~1,
  p.~6948, 2017.

\bibitem{Goverdovsky2015}
V.~Goverdovsky, D.~Looney, P.~Kidmose, C.~Papavassiliou, and D.~P. Mandic,
  ``{Co-located multimodal sensing: A next generation solution for wearable
  health},'' {\em IEEE Sensors Journal}, vol.~15, no.~1, pp.~138--145, 2015.

\bibitem{Goverdovsky2016}
V.~Goverdovsky, D.~Looney, P.~Kidmose, and D.~P. Mandic, ``{In-ear EEG from
  viscoelastic generic earpieces: Robust and unobtrusive 24/7 monitoring},''
  {\em IEEE Sensors Journal}, vol.~16, no.~1, pp.~271--277, 2016.

\bibitem{Curran2016}
M.~T. Curran, J.-k. Yang, N.~Merrill, and J.~Chuang, ``{Passthoughts
  authentication with low cost earEEG},'' in {\em Proceedings of the Annual
  International Conference of the IEEE Engineering in Medicine and Biology
  Society, EMBS}, pp.~1979--1982, 2016.

\bibitem{Neuper2005}
C.~Neuper, R.~H. Grabner, A.~Fink, and A.~C. Neubauer, ``{Long-term stability
  and consistency of EEG event-related (de-)synchronization across different
  cognitive tasks},'' {\em Clinical Neurophysiology}, vol.~116, no.~7,
  pp.~1681--1694, 2005.

\bibitem{hayes1996statistical}
M.~H. Hayes, ``{Statistical digital signal processing and modeling},'' 1996.

\bibitem{murphy2012machine}
K.~P. Murphy, {\em {Machine learning: a probabilistic perspective}}.
\newblock MIT press, 2012.

\bibitem{Chang2011}
C.-C. Chang and C.-J. Lin, ``{LIBSVM: A library for support vector machines},''
  {\em ACM Transactions on Intelligent Systems and Technology (TIST)}, vol.~2,
  no.~3, p.~27, 2011.

\bibitem{Landis2008}
J.~R. Landis and G.~G. Koch, ``{The measurement of observer agreement for
  categorical data},'' {\em Biometrics}, vol.~33, no.~1, pp.~159--174, 1977.

\bibitem{Silber2007}
M.~H. Silber, S.~Ancoli-Israel, M.~H. Bonnet, S.~Chokroverty, M.~M.
  Grigg-Damberger, M.~Hirshkowitz, S.~Kapen, S.~A. Keenan, M.~H. Kryger,
  T.~Penzel, M.~R. Pressman, and C.~Iber, ``{The visual scoring of sleep in
  adults},'' {\em Journal of Clinical Sleep Medicine}, vol.~3, no.~2,
  pp.~121--131, 2007.

\end{thebibliography}
}

\end{document}